\documentclass{caps}
\usepackage{epsfig}
\begin{document}

\pagenumbering{roman}
\cleardoublepage

\pagenumbering{arabic}

\setcounter{chapter}{2}

\setcounter{table}{0}

\pagenumbering{arabic}

\author[Strohmayer and Bildsten]{Tod Strohmayer\\Laboratory 
for High Energy Astrophysics\\NASA's Goddard Space Flight Center, Greenbelt, 
MD 20771 \and Lars Bildsten\\Kavli Institute for Theoretical Physics and
Department of Physics \\ University of California, Santa Barbara, CA 93106}
\chapter{New Views of Thermonuclear Bursts}

\section{Introduction}

Many accreting neutron stars erupt in spectacular thermonuclear
conflagrations every few hours to days. These events, known as Type I
X-ray bursts, or simply X-ray bursts, are the subject of our
review. Since the last review of X-ray burst phenomenology was written
(Lewin, van Paradijs \& Taam 1993; hereafter LVT), powerful new
X-ray observatories, the  Rossi X-ray Timing Explorer (RXTE), the
Italian - Dutch BeppoSAX mission, XMM-Newton  and Chandra 
have enabled the discovery of entirely new phenomena associated with 
thermonuclear burning on neutron stars. Some of these new findings include: 
(i) the discovery of millisecond (300 - 600 Hz) oscillations during bursts, 
so called ``burst oscillations'', (ii) a new regime of nuclear burning on
neutron stars which manifests itself through the generation of hours
long flares about once a decade, now referred to as ``superbursts'',
(iii) discoveries of bursts from low accretion rate neutron stars, and
(iv) new evidence for discrete spectral features from bursting neutron
stars.

It is perhaps surprising that nuclear physics plays such a prominent
role in the phenomenology of an accreting neutron star, as the
gravitational energy released per accreted baryon (of mass $m_p$) is
$GMm_p/R\approx 200 \ {\rm MeV}$ is so much larger than the nuclear
energy released by fusion ($\approx 5 \ {\rm MeV}$ when a solar mix
goes to heavy elements). Indeed, if the accreted fuel was burned
at the rate of accretion, any evidence of nuclear physics would be
swamped by the light from released gravitational energy. The only way
the nuclear energy can be seen is when the fuel is stored for a
long period and then burns rapidly (as in Type I bursts and
Superbursts).

The advances in millisecond timing were enabled by RXTE's combination
of large X-ray collecting area, sub-millisecond time resolution and a
state-of-the-art data system and telemetry capacity.  RXTE's
instruments are also unfettered; they can look where the action is for
longer, more often, and more quickly than any previous X-ray
observatory. The Wide Field Cameras (WFC) on BeppoSAX (Jager et
al. 1997), in combination with the All Sky Monitor (ASM) onboard RXTE
(Levine et al. 1996), have provided an unprecedented, long term view
of the X-ray sky that opened up new discovery space for rare
events. Their capabilities led to the discovery of superbursts as a
new manifestation of thermonuclear burning on neutron stars, as well
as the detection of rare bursts from neutron stars accreting at very
low rates. Indeed, WFC observations alone have led to the discovery of
about 20 new bursters (see in 't Zand 2001).

In addition to these new results, recent discoveries have provided
answers to some long standing questions concerning the sources of
thermonuclear bursts, the neutron star low mass X-ray binaries
(LMXB). These include; (i) the discovery of accreting millisecond
pulsars in three LMXB systems, confirming that neutron stars are
spun-up to millisecond periods by accretion, (ii) the discovery of
sub-millisecond variability from many neutron star LMXBs (see the
review by van der Klis in Chapter 2) and (iii) detections of several
bursting neutron star transients (e.g. Aql X-1, Cen X-4) in quiescence
with the increased sensitivity afforded by Chandra and
XMM/Newton. These data have provided important new information on the
thermal state of neutron stars which strongly influences their
bursting behavior (see Chakrabarty \& Psaltis in Chapter 1).

Our review focuses on the new phenomena and insights they
provide about neutron stars and thermonuclear burning on
them. These discoveries have stimulated much new theoretical thinking.
In particular, the discovery of millisecond oscillations during bursts
has refocused attention on the need to understand the multidimensional
nature of thermonuclear flame propagation, that is, how burning is
initiated and how it spreads around the neutron star. An understanding
of flame propagation, in turn, requires an understanding of how
thermonuclear burning influences the dynamics of the accreted
layers. To understand superbursts, theorists have been forced
to think about what happens at much greater depths in the accreted
layers of neutron stars, and about the detailed composition of the
``ashes'' of H/He burning. Finally, observations of bursts from
sources with very low mass accretion rates are forcing theorists to
confront some physical processes, for example, element diffusion,
which have not been considered in earlier work.

Space constraints do not allow us to present an exhaustive review of
all X-ray burst phenomenology and theory. For example, we will not describe 
in detail the Type II (accretion instability) burst phenomena now observed from
two sources; the famous Rapid Burster (see for example, LVT); and the
Bursting Pulsar (GRO J1744--28, see Kouveliotou et al. 1996; Giles et
al. 1996). Although new observations of Type II bursters have been
obtained since the last reviews were written, including the discovery
of GRO J1744-28, only the 2nd known Type II burster, they have
generally not provided new qualitative insights on the accretion
instability. Some of the more recent observations of these sources,
and our current understanding of them, are described elsewhere in this
volume (see the contributions by Lewin \& Verbunt; Chakrabarty \&
Psaltis; and van der Klis).

Despite its energetic disadvantage, we are now confident that sudden
nuclear energy release powers bursts and, as we emphasize here,
fleshing out the details of the thermonuclear flash model in the
context of the observations is probing both neutron star structure and
fundamental physics. We try not to segregate our discussions into
separate observational and theoretical pieces, rather, we
integrate the theory and observations as they relate to particular
phenomena as much as possible. In that sense, this is an incomplete
review of the state of the theoretical research in this field.
 
We begin in \S 2 with a brief theoretical introduction to the relevant
physics of thermonuclear burning on neutron stars and how it accounts
for the gross properties of bursts.  We then briefly review the
observational characteristics of bursts and sources. 
This will lay out the fundamentals and topics important for our
discussions of the new results. Here we also include some discussion on
new results from burst spectroscopy. We refer the reader to Bildsten (1998)
for relevant theoretical details.

We then move to the new discoveries. In \S 4 we introduce
millisecond variability during bursts (``burst oscillations''), and
theoretical implications motivated by the observations. In \S 5 we
describe the new observations of superbursts and their theoretical
implications.  We close in \S 6 with a summary and some comments on
the future prospects of burst research.

\section{The physics of hydrogen-helium burning}
 
At the core of the observed phenomena is the ``thin shell''
instability discovered theoretically by Schwarzschild \& H$\ddot {\rm
a}$rm (1965) in the helium shell residing above the carbon/oxygen core
during the asymptotic giant branch phase of stellar evolution. The
driver of this instability is a nuclear energy generation rate that is
more temperature sensitive than radiative cooling and is confined to a thin 
shell. Hansen \& Van Horn
(1975) showed that burning of the accumulated hydrogen and helium on a
neutron star also occurs in radially thin shells that were susceptible
to the same instability. Soon thereafter, type I X-ray bursts
from LMXBs were independently discovered by Grindlay et al. (1976) and
Belian, Conner \& Evans (1976), and were quickly associated (Woosley
\& Taam 1976; Maraschi \& Cavaliere 1977; Joss 1977, 1978; Lamb \& Lamb
1978) with Hansen \& Van Horn's (1975) instability.  For a brief
historical overview see LVT (1993).

The successful association of thermonuclear instabilities with X-ray
bursts made a nice picture of a recurrent cycle that consists of fuel
accumulation for several hours to days followed by a thermonuclear
runaway that burns the fuel in $\sim 10-100$ seconds. It also secured
the identification of the accreting objects as neutron stars (NS). The
mass donors---the ultimate source of the thermonuclear fuel---are
typically old, Population II objects or in some cases, degenerate
helium or perhaps carbon/oxygen white dwarfs (Rappaport, Joss \&
Webbink 1982). The accreted composition is important, as the nuclear
ashes and burst properties depend on the accreted mix of light
elements.  Unfortunately, in most cases, we have little information on
the composition of the accreted fuel. Of the approximately 160 known
LMXB's about 70 are observed to produce bursts (see Liu, van Paradijs
\& van den Heuvel 2001; Chakrabarty \& Psaltis in Chapter 1).

A fundamental physical reason for studying the neutron star example of
a thin shell instability is that the timescales are observationally
accessible.  Hundreds of bursts have been seen from some neutron
stars, and study of the burst dependence on accretion rate can be
undertaken.  Such an exercise is rare to impossible in the other
astrophysical site where the instability is observable: the explosion
of hydrogen on an accreting white dwarf as a classical nova. For a more
detailed theoretical discussion see Bildsten (1998).

\subsection{Nuclear burning during accumulation and ignition}

After the accreted hydrogen and helium has become part of the NS, it
undergoes hydrostatic compression as new material is piled on. The
fresh fuel reaches ignition densities and temperatures within a few
hours to days. The resulting compression rate depends on the accretion
rate per unit area, $\dot m\equiv \dot M/A_{acc}$, where $A_{acc}$ is
the area covered by fresh material. We will sometimes quote numbers
for both $\dot m$ and $\dot M$. When we give $\dot m$, we have assumed
$A_{acc}=4\pi R^2\approx 1.2\times 10^{13} \ {\rm cm^2}$. The time it
takes for heat transport to cool the deep envelope (what we call the
thermal time) is only ten seconds at the ignition location, where the
pressure is $P\approx 10^{22}-10^{23} \ {\rm erg \ cm^{-3}}$. This is
so much shorter than the time to accumulate the material (hours to
days) that the compression is far from adiabatic.

The temperature exceeds $10^7$ K in most of the accumulating
atmosphere, so that hydrogen burns via the CNO cycle and the proton -
proton (pp) cycle can be neglected. At high temperatures ($T>8\times
10^7 \ {\rm K}$), the timescale for proton captures becomes shorter
than the subsequent $\beta$ decay lifetimes, even for the slowest
$^{14}$N(p,$\gamma$)$^{15}$O reaction. The hydrogen then burns in the
``hot'' CNO cycle of Fowler \& Hoyle (1965)
\begin{equation} 
^{12}{\rm C}(p,\gamma)^{13}{\rm N}(p,\gamma)^{14}{\rm
O}(\beta^+)^{14}{\rm N}(p,\gamma)^{15}{\rm O}(\beta^+)^{15} {\rm
N}(p,\alpha)^{12} {\rm C}, 
\end{equation} 
and is limited to $5.8\times 10^{15} Z_{\rm CNO} {\rm \ ergs \ 
g^{-1} \ s^{-1}}$, where $Z_{\rm CNO}$ is
the mass fraction of CNO. The hydrogen burns this way in the
accumulating phase when, 
\begin{equation}
\dot m> 900 \ {\rm g \ cm^{-2} \ s^{-1}} (Z_{CNO}/0.01)^{1/2} \; ,
\end{equation}
and is thermally stable. The amount of time it
takes to burn the hydrogen is $\approx (10^3/Z_{\rm CNO}) \ {\rm s}$,
or about one day for solar metallicities. This time is even longer if
the donor star has a low metal content or if there is substantial 
spallation of the incident CNO elements as discussed by Bildsten,
Salpeter and Wasserman (1992). For lower $\dot m$'s, the
hydrogen burning is thermally unstable and can trigger Type I bursts.

The slow hydrogen burning during accumulation allows for a unique
burning regime at high $\dot m$'s.  This simultaneous H/He burning
occurs when
\begin{equation}
\dot m > 2\times 10^{3} \ {\rm g \ cm^{-2} \ s^{-1}}
(Z_{CNO}/0.01)^{13/18} \; , 
\end{equation}
(Bildsten 1998; Cumming \& Bildsten 2000), as
at these high rates the fluid is compressed to helium ignition
conditions before the hydrogen is completely burned (Lamb \& Lamb
1978; Taam \& Picklum 1978). The strong temperature dependence of the
helium burning rate (and lack of any weak interactions) leads to a
thin-shell instability for temperatures $T< 5 \times 10^8 \ {\rm K}$
and causes the burst. The critical condition of thin burning shells
($h\ll R$) is true before burning and remains so during the flash.
Stable burning sets in at higher $\dot m$'s (comparable to the
Eddington limit; Paczynski 1983; Bildsten 1998, Narayan \& Heyl 2002) 
when the helium burning temperature sensitivity
finally becomes weaker than the cooling rate's sensitivity (Ayasli \&
Joss 1982; Taam, Woosley \& Lamb 1996). This is consistent with the 
absence of bursts from high-field X-ray pulsars, which channel the 
accretion flow onto a small-area polar cap, and thus achieve a high
local (and stabilizing) $\dot m$ (Joss \& Li 1980; Bildsten \& Brown 1997). 

For solar metallicities, there is a narrow window of $\dot m$'s where
the hydrogen is completely burned before the helium ignites. In this
case, a pure helium shell accumulates underneath the hydrogen-burning
shell until conditions are reached for ignition of the pure helium
layer. The recurrence times of these bursts must be longer than the
time to burn all of the hydrogen, so pure helium flashes should have
recurrence times in excess of a day and $\alpha\approx 200$ ($\alpha$
is the ratio of the time-averaged persistent to burst luminosity). To
summarize, in order of increasing $\dot m$, the regimes of unstable
burning we expect from NSs accreting at sub-Eddington rates ($\dot
m<10^5 \ {\rm g \ cm^{-2} \ s^{-1}}$) are (Fujimoto, Hanawa \& Miyaji
1981; Fushiki \& Lamb 1987; Cumming \& Bildsten 2000):
\begin{enumerate}

\item{Mixed hydrogen and helium burning triggered by thermally 
unstable hydrogen ignition for $\dot m < 900 \ {\rm g \ cm^{-2} \
s^{-1}}$ ($\dot M< 2\times 10^{-10} M_\odot \ {\rm yr^{-1}}$).} 

\item{Pure helium shell ignition for 
$900 \ {\rm g \ cm^{-2} \ s^{-1}}< \dot m < \ 2 \times 10^3 
\ {\rm g \ cm^{-2} \ s^{-1}}$ following completion of hydrogen burning.}

\item{Mixed hydrogen and helium burning triggered by thermally 
unstable helium ignition for $\dot m > 2 \times 10^3 \ {\rm g \
cm^{-2} \ s^{-1}}$ ($\dot M>4.4\times 10^{-10} M_\odot \ {\rm
yr^{-1}}$).} 
\end{enumerate} 

\noindent 
The transition $\dot m$'s are for $Z_{CNO}\approx 0.01$. Reducing
$Z_{CNO}$ lowers the transition accretion rates and substantially 
narrows the $\dot m$ range for pure helium ignition. 
Another effect critical to the burning is the amount of heat flux
coming through the burning layer from deeper parts of the NS ocean
(Ayasli \& Joss 1982, Fushiki \& Lamb 1987). The
current theoretical estimates (Brown, Bildsten \& Rutledge 1998; Brown
2000; Colpi et al. 2001) are that between 10 and 100\% of the heat
released via pycnonuclear reactions in the deep crust (Haensel \&
Zdunik 1990) emerges from the surface, proving most important to the
burst properties of pure helium accretors such as 4U 1820--30 
(Bildsten 1995, Strohmayer \& Brown 2002). 

\begin{figure*}
 \centering \hspace{-0.23cm}
 \epsfig{file=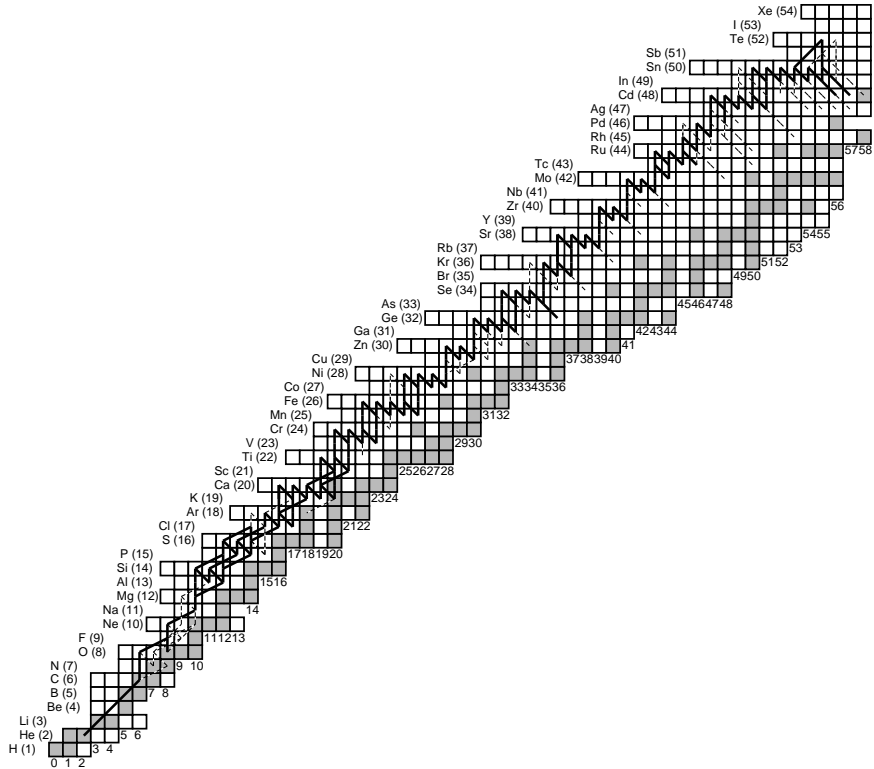,width=0.9\linewidth}
\vskip 0.1in \caption{Schematic showing the dominant pathways of the
nuclear reaction flows during the rp process.  Elements far beyond
$^{56}$Fe can easily be reached. Filled squares denote stable nuclides
(after Schatz et al. 2001).}
\label{figure1}
\end{figure*}

\subsection{Burning during the bursts: the rp-process}

We now briefly discuss what happens as the thermal instability
develops into a burst and what observational differences are to be
expected between a pure helium ignition and a mixed hydrogen/helium
ignition. The flash occurs at fixed pressure, and the increasing
temperature eventually allows the radiation pressure to dominate.  For
a typical ignition column of $2\times 10^8 \ {\rm g \ cm^{-2}}$, the
pressure is $P=gy\approx 4\times 10^{22} \ {\rm ergs \ cm^{-3}}$, so
$aT_{max}^4/3\approx P$ gives $T_{max}\approx 2\times 10^9 {\rm \
K}$. For pure helium flashes, the fuel burns rapidly (since there are
no slow weak interactions) and the local Eddington limit is often
exceeded. These conditions lead to photospheric radius expansion (PRE)
bursts with durations, set mostly by the time it takes the heat to
escape, of order $5-10$ seconds.

When hydrogen and helium are both present, the temperatures reached
during the thermal instability can easily produce elements far beyond
the iron group (Hanawa et al. 1983; Wallace \& Woosley 1984; Hanawa \&
Fujimoto 1984; Koike et al. 1999; Schatz et al. 1999, 2001) via the
rapid-proton (rp) process of Wallace \& Woosley (1981). This burning
starts after the triggering helium flash heats the gas to high enough
temperatures to allow the ``breakout'' reactions 
$^{15}{\rm O}(\alpha,\gamma)^{19}{\rm
Ne}$ and $^{18}{\rm Ne}(\alpha,p)^{21}{\rm Na}$ ($^{18}{\rm
Ne}$ is made in the chain $^{14}{\rm
O}(\alpha,p)^{17}{\rm F}(p,\gamma)^{18}{\rm Ne}(\beta^+)^{18}
{\rm F}(p,\alpha)^{15}{\rm O}$ triggered by the high temperatures) 
to proceed faster than the $\beta$ decays.
This takes these catalysts out of the CNO cycle loop, where they 
subsequently burn hydrogen via the rp process: a series of successive 
proton captures and $\beta$ decays.

Figure 3.1 shows the dominant path of the nuclei
as they move up the proton-rich side of the valley of stability (much
like the r-process which occurs by neutron captures on the neutron
rich side) more or less limited by the $\beta$-decay
rates. Theoretical work shows that the end-point of this
time-dependent burning is far beyond iron (Hanawa \& Fujimoto 1984,
Schatz et al. 1999, 2001, Koike et al. 1999) and is set by either the
complete burning of the hydrogen or reaching the closed SnSbTe
cycle found by Schatz et al. (2001). When hydrogen is 
exhausted prior to reaching the SnSbTe cycle, a rough estimate of the 
rp-process end-point is made by merely accounting for the ratios of seed 
nuclei to hydrogen. This produces a large range of heavy nuclei (Schatz et
al. 1999, 2001).  Schatz et al. (1999) also showed an additional
important point of nucleosynthesis, which is that the hydrogen
burns out before the helium is completely burned.  Thus, the carbon
made during late helium burning remains as carbon since there are no
protons available for it to capture. As we will show in Section 5,
this remaining carbon is the apparent fuel for the recently discovered
superbursts. Observationally, the long series of $\beta$ decays during
the rp-process releases energy for at least 150 seconds after the
burst has started. We thus expect a mixed hydrogen/helium burst to
last much longer than a pure helium burst.

\subsection{Mixed H/He bursts from GS 1826--238} 

\begin{figure*}[tbp]
 \centering
 \epsfig{file=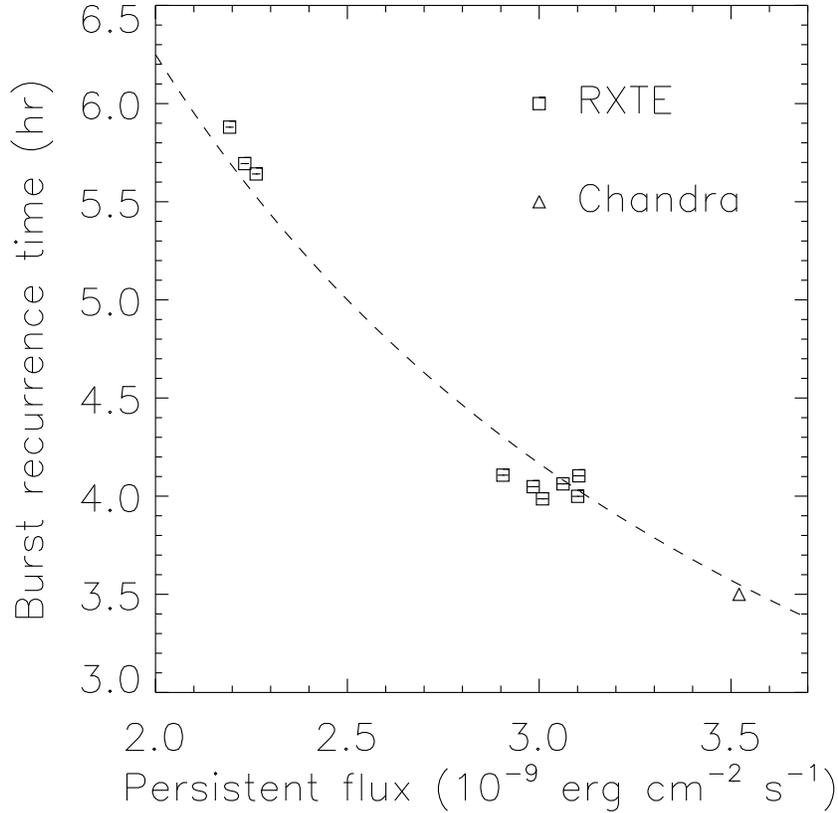,width=0.85\linewidth}
\vskip 0.1in \caption{
Variation of the burst recurrence time for GS 1826--238 as a function
of the persistent flux from RXTE measurements between 1997--2000 
(squares) and a more recent {\it Chandra}/RXTE measurement on 2002 July 29. 
Horizontal error bars indicate the $1\sigma$ errors. The dashed line
is the trend expected if the burst recurrence time is $\propto \dot M^{-1}$ 
(after Galloway et al. 2003). See also Figure 3 in Cornelisse et al. (2003).}
\label{figure2}
\end{figure*}

We discuss shortly that most bursters do not precisely match the
theory presented above in a simple way, and initial indications are
that this could be due to variations in the area on which new fuel
accretes (Bildsten 2000). However, there are times when bursters behave in a
limit cycle manner, with bursts occurring periodically (e.g. Robinson
\& Young 1997) as $\dot m$ stays at a fixed value and we discuss one
beautiful example here: the Type I burster GS 1826--238. Ubertini et
al. (1999) found 70 bursts over a 2.5 year monitoring baseline with
the BeppoSAX/WFC. The quasi-periodic recurrence time was $5.76\pm
0.62$ hours. Cocchi et al.'s (2000) later analysis within observing
seasons found that in 1997 and 1998 the recurrence times were even
more clock-like, $5.92\pm 0.07$ hours and $5.58\pm 0.09$ hours. The
persistent flux during this bursting period was $F_x\approx 2\times
10^{-9} \ {\rm erg \ cm^{-2} \ s^{-1}}$ (Ubertini et al. 1999, in 't
Zand et al. 1999, Kong et al. 2000).
Since then, the persistent flux has risen, and remarkably, the burst
recurrence time has shortened $\propto \dot M^{-1}$ (see Figure 3.2, 
Galloway et al. 2003), implying that the accumulated mass is the same 
as $\dot M$ changes. 

\begin{figure*}[tbp]
 \centering
 \epsfig{file=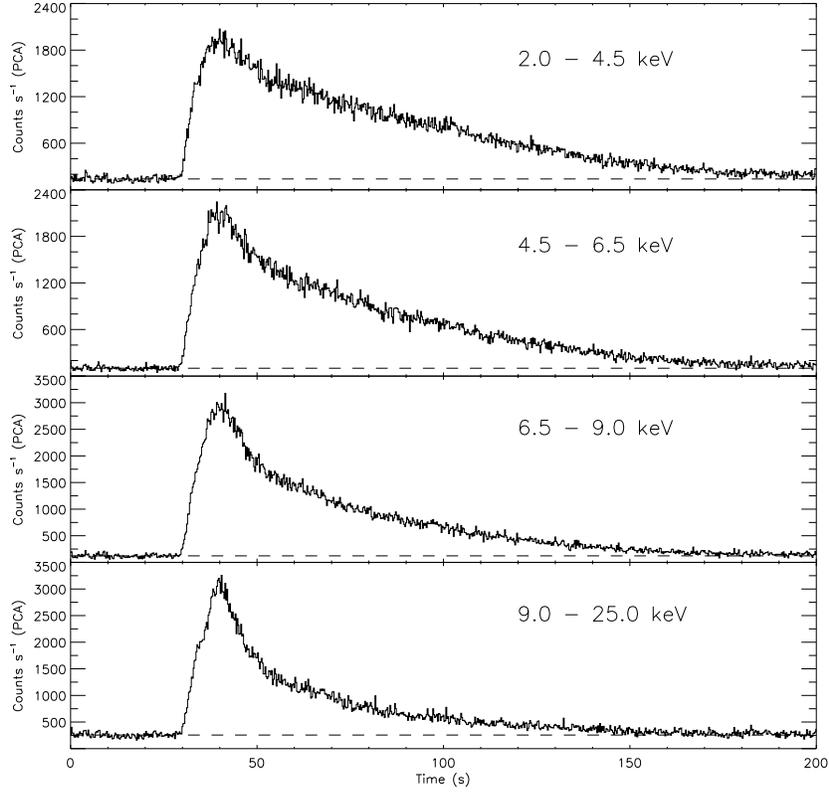,width=0.85\linewidth}
\vskip 0.1in \caption{An X-ray burst from GS 1826--238 seen with the
RXTE/PCA.  The burst is shown in four different energy bands. The long
duration is indicative of the delayed energy release from the rapid
proton (rp) process. The dashed line marks the preburst flux level
(see also Kong et al. 2000).}
\label{figure3}
\end{figure*}

From the observed persistent flux and the measured ratio of
$R_{app}/d$ (about 10 km at 10 kpc; Galloway et al. 2003) the local
accretion rate is $\dot m\approx 10^4 \ {\rm g \ cm^{-2} \ s^{-1}}$
when $F_x\approx 2\times 10^{-9} \ {\rm erg \ cm^{-2} \ s^{-1}}$. This
is very safely in the mixed H/He burning regime. The estimated $\dot
m$ gives an accumulated column on the NS prior to the burst of
$2\times 10^8 \ {\rm g \ cm^{-2}}$, just what is expected from theory
(see Table 2 of Cumming \& Bildsten 2000). These periodic bursts allow
for a very secure measurement of $\alpha\approx 40$ (Galloway et
al. 2003), implying a nuclear energy release of $5 \ {\rm MeV}$ per
accreted nucleon for a $1.4 \ M_\odot$, 10 km NS. Energy releases this
large can only come about via hydrogen burning and the long ($>100 $ s) 
duration of the bursts are consistent with the expected energy
release from the rp-process.  Figure 3.3 shows the time profile of a
burst seen with the Proportional Counter Array (PCA) on RXTE (see also
Kong et al. 2000). These data provide an important confirmation of the
delayed energy release expected when hydrogen is burning via an
rp-process. The resemblance of these profiles to theoretical results
of Hanawa \& Fujimoto (1984) and Schatz et al. (2001) is striking.

\section{Observational overview of bursts}

In this section we give a brief overview of important observational
characteristics of bursts.  We will emphasize recent work and specific
topics which are relevant in the context of the new discoveries to be
described in later sections. We refer to earlier reviews for some of
the details we must omit.

Bursts are most commonly observed from the ``atoll'' subclass of LMXBs
with luminosities above $\approx 10^{36}$ ergs s$^{-1}$ (Hasinger \&
van der Klis 1989; LVT). These systems are accreting at intermediate
rates of between $0.01 - 0.3 \ \dot M_{Edd}$, which is roughly
consistent with the accretion rate range of the thermonuclear
instabilities described earlier.  Interestingly, observations with the
WFC on BeppoSAX have recently discovered bursts from 10 LMXBs with
persistent X-ray luminosities significantly less than $10^{36}$ ergs
s$^{-1}$ (see Cornelisse et al. 2002a; Cornelisse et al.  2002b;
Cocchi et al. 2001; \S 3.3.4). These recently discovered bursts probe the 
mass accretion rate dependence of thermonuclear burning in a previously 
unexplored regime. For example, physical processes such as element diffusion, 
which are not relevant to most bursters, likely become important at such low 
accretion rates (Wallace, Woosley \& Weaver 1982). 
We will say some more about these bursts shortly.

\subsection{Burst profiles and spectra}

Though the time profiles of X-ray bursts are diverse, they do share
several characteristic features.  (i) Burst rise times are shorter
than their decay times. Rise times are typically $< 2$ s, but in some
cases can be as long as 10 s. Excluding superbursts, which we discuss
later, burst decay times range from about 10 s to several
minutes, with most bursts having 10 - 20 s decays.  (ii) Burst
profiles are shorter at higher energies. This is a direct result of
cooling of the neutron star surface with time. (iii) Burst profiles
are generally smooth, showing an exponential or exponential-like
intensity decay.

Swank et al. (1977) and Hoffman, Lewin \& Doty (1977) showed that
bursts have thermal (blackbody) spectra.  The radius of a blackbody
emitting a flux, $F_{bol}$, at temperature, $T_{bb}$, is; $R_{bb} = d
( F_{bol} / \sigma T_{bb}^4 )^{1/2}$, where $d$ is the source
distance. Measurement of the bolometric fluxes and blackbody
temperatures can then be used to infer radii, if the distance is
known.  Radii inferred in this manner are typically in the range of
$\sim 10$ km, consistent with cooling of an object having the
theoretical size of a neutron star.

Although burst spectra are observationally well described by the
Planck function, theoretically they should be harder than a blackbody
at the effective temperature of the atmosphere (see London, Howard \&
Taam 1984, 1986; Ebisuzaki \& Nakamura 1988; Madej 1991; Titarchuk 1994). 
This occurs because electron (Compton) scattering is an important opacity 
source in a neutron star atmosphere.  Direct evidence for this effect comes
from the peak blackbody temperatures ($kT_{bb} > 3$ keV) of some
bursts, which are significantly higher than the Eddington effective
temperature for reasonable neutron star models (see London, Howard \&
Taam 1984; 1986).

\subsection{Photospheric radius expansion bursts}

In bright bursts the local X-ray luminosity in the atmosphere may reach the
Eddington limit
\begin{equation}
L_{Edd} = \left ( 4\pi  c G M / \kappa \right ) \left ( 1 - 2GM/c^2 R 
\right )^{-1/2} \ = \ 4\pi R^2 \sigma T_{eff}^4,
\end{equation}
where $M$, $R$, and $\kappa$ are the neutron star mass, radius and atmospheric
opacity, respectively, and the photospheric layers can be lifted off the
neutron star surface by radiation pressure. Note that $L_{Edd}$ depends
on the composition of the accreted atmosphere through the opacity $\kappa$. 
In these bursts the blackbody temperature decreases while the inferred 
blackbody radius simultaneously increases.  This all happens while the 
total X-ray flux stays approximately constant.  These bursts are called 
photospheric radius expansion (PRE) bursts. The moment when the photosphere 
falls back to the neutron star surface (when the temperature is highest) is
called ``touchdown.''  Theoretical work indicates that in such bursts
the X-ray flux stays within a few percent of the Eddington limit, and
the excess energy is efficiently transferred into kinetic energy of
the outflow (Hanawa \& Sugimoto 1982; Paczynski \& Anderson 1986; Joss \& 
Melia 1987; Titarchuk 1994; Shaposhnikov \& Titarchuk 2002). Ebisuzaki \& 
Nakamura (1988) found evidence for photospheric composition variations when 
comparing fainter bursts with PRE bursts from the LMXBs 4U 1608--52 and 
4U 1636--53. They found differences in the luminosity -- color temperature 
relation between these two classes which they attributed to atmospheric 
composition changes. They suggested that the luminosity -- color temperature 
relation for faint bursts and PRE bursts could be explained by hydrogen rich 
and hydrogen poor atmospheres, respectively. They further hypothesized that 
the hydrogen rich envelope is ejected during bright PRE bursts, and thus the 
atmospheric composition is pure helium in these cases. 

The amount of photospheric uplift can vary dramatically from burst to burst 
and amongst different sources. In the most powerful bursts the expansion can 
be large enough to shift the effective temperature of the photosphere 
entirely below the X-ray band (See Lewin, Vacca \& Basinska 1984; 
Strohmayer \& Brown 2002). Such events show ``precursors'' separated from 
the main part of the burst by the cooling of the photosphere. Other less 
powerful bursts may show double peaked profiles as only a portion of the 
flux is shifted out of the X-ray band by the expansion. Figure 3.4 shows 
several examples of bursts, both with and without PRE, observed with the 
RXTE/PCA from the LMXB 4U 1728--34.

Since the Eddington luminosity should impose an upper limit to burst
fluxes, it was suggested early on that bursts might provide a
``standard candle'' (van Paradijs 1978), and that they could be used
as distance indicators.  Van Paradijs (1981) recognized that globular
cluster bursters, with independently known distances, would provide an
important test of this idea.  Subsequent researchers concluded that
the brightest (PRE) bursts likely represent a true limiting
luminosity, but some uncertainties remained (Lewin 1982; Basinska et
al. 1984). Recently Kuulkers et al. (2002a) have reexamined this issue
using the extensive sets of bursts observed with RXTE and BeppoSAX/WFC,
and a uniform set of globular cluster distances. They conclude that
the radius expansion bursts can indeed be regarded as standard candles
to within about 15 \%, and derive a critical luminosity of about $3.8
\times 10^{38}$ erg s$^{-1}$, consistent with the Eddington limit for
hydrogen-poor matter from a neutron star. Since we do not expect that all
LMXBs are pure helium accretors, the fact that PRE bursts from the globular
cluster sources have approximately the same peak fluxes provides additional
support to the idea that in these bursts the hydrogen is blown off in a wind.  

Galloway et al. (2002) have recently examined the distribution of peak 
fluxes of bursts from 4U 1728--34. They find that the peak fluxes of PRE 
bursts are not constant, but show variations of $\approx 44 \%$.  However, 
they find this variation is correlated with the source spectral state just
prior to the bursts, and suggest it may be caused by reprocessing from
a precessing, warped disk. Upon removing the correlation they obtain a
3 \% variation. They use this result to argue that during radius expansion
episodes the emission from the photosphere is largely isotropic. Assuming the 
peak flux is the Eddington limit appropriate for pure helium they derive a 
source distance of 5.2 - 5.6 kpc. Smale (1998) and Kuulkers \& van der Klis 
(2000) used radius expansion bursts observed with RXTE from Cyg X-2 and 
GX 3+1, respectively, to place new constraints on the distances to these 
sources.  

\begin{figure*}
 \centering
 \epsfig{file=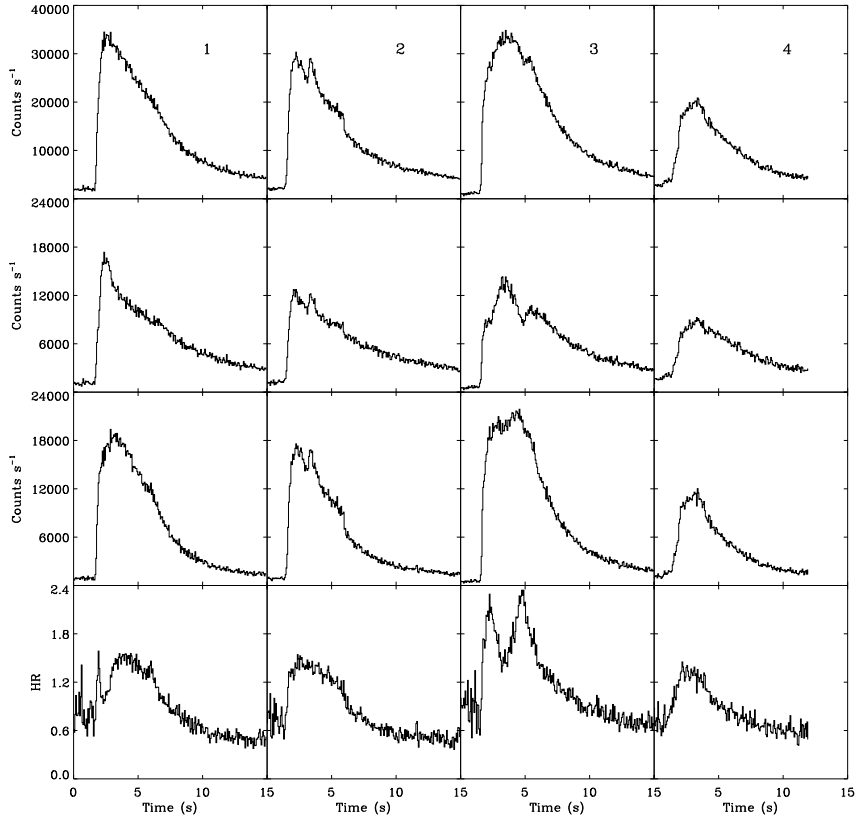,width=0.8\linewidth}
\vskip 0.1in \caption{A sample of four X-ray bursts from the LMXB 4U
1728--34 as observed with the RXTE/PCA. Each sequence shows, from top
to bottom, the total 2 - 60 keV countrate, the 2 - 6 kev countrate,
the 6 - 30 keV countrate, and the hardness ratio (6 - 30 keV) / (2 - 6
keV). Bursts 1 and 3 show clear evidence for PRE based on the hardness
ratio evolution.}  
\label{figure4}
\end{figure*}

In principle, observations of PRE bursts provide a means to infer the
masses and radii of neutron stars, quantities which have profound
implications for our understanding of the equation of state of dense
matter (Lattimer \& Prakash 2001).  Here we will only
outline the main ideas, and we refer the reader to LVT for a more
detailed discussion. As noted above, the Eddington luminosity at the 
surface of a neutron star depends only on the stellar mass, $M$, radius, 
$R$, and the composition of the atmosphere.
The above expression (see Equation 3.4) shows that a measurement of the 
effective temperature, $T_{\rm eff}$, when the luminosity is Eddington limited,
gives a constraint on the mass and radius of the neutron star. Indeed,
this constraint is independent of the distance to the source. In
practice, however, it is model dependent because the
observed color temperature (from, say, a blackbody spectral fit),
must be corrected to an effective temperature using an atmosphere
model. Typically there are no independent constraints on the
atmospheric composition, and so the correct model is uncertain.
Measurement of the variation of the Eddington luminosity caused by
expansion of the photosphere, and hence a change in the gravitational
redshift factor, can also, in principle, be used to determine the
gravitational redshift from the neutron star surface (see for example,
Damen et al. 1990; van Paradijs et al. 1990; LVT). If the distance to
the source is known, then a measure of the burst flux when it is
Eddington limited will constrain a slightly different function of $M$
and $R$. Sources with independent distance constraints, such as the
globular cluster bursters, can provide, in principle, the best
constraints (see Kuulkers 2002a).

Many researchers have used these and related methods to try and
constrain neutron star masses and radii (see Fujimoto \& Taam 1986;
Ebisuzaki 1987; Sztajno et al. 1987; van Paradijs \& Lewin 1987;
Chevalier \& Ilovaisky 1990; Kaminker et al. 1990; Damen et al. 1990;
van Paradijs et al. 1990; Haberl \& Titarchuk 1995). In general, the
constraints are consistent with a range of neutron star mass - radius
relations, but because of systematic uncertainties in the spectra,
composition, and burst flux isotropy, they are not generally precise
enough to unambiguously constrain the neutron star equation of state.
Such efforts have continued using the higher signal to noise data from
RXTE (see for example Smale 2001; Strohmayer et al. 1998; Titarchuk \&
Shaposhnikov 2002), however, systematic uncertainties remain.

\subsection{Recent progress in burst spectroscopy}

The most reliable way to overcome the systematic uncertainties in the
interpretation of continuum spectra from bursts is to detect line
features from the neutron star surface, and thereby obtain a direct
measurement of the gravitational redshift, $1 + z = (1 - 2GM/c^2 R
)^{-1/2}$. Waki et al. (1984), and Nakamura, Inoue \& Tanaka (1988)
reported absorption lines at 4.1 keV in TENMA data from the LMXBs 4U
1636--53, and 4U 1608--52. Magnier et al (1989) found a similar feature
from 4U 1747--214 using EXOSAT data. Waki et al. (1984) identified the
line with the helium-like iron Ly$\alpha$ transition at 6.7 keV, but
gravitationally redshifted from the neutron star surface.  The high
implied equivalent widths (hundreds of eV), however, led others to
suggest an origin outside the neutron star atmosphere, perhaps in the
accretion flow (see Day, Fabian \& Ross 1992; Foster, Fabian \& Ross
1987; Pinto, Taam \& Laming 1991). More recent observations, for
example, with ASCA, RXTE and BeppoSAX, have generally not confirmed
the presence of such lines in burst spectra. The interpretation, and
reality, of these features has therefore remained controversial.

In spite of this controversy, reports of deviations from blackbody 
spectra during some bursts continue to appear. Kuulkers et al. (2002b)
found systematic deviations from blackbody spectra in RXTE data of
bursts from the high accretion rate Z source GX 17+2. The residuals
were most significant during the radius expansion and contraction
phases. These variations are quite similar to those described by van
Paradijs et al. (1990) for a PRE burst from 4U 2129+11, and, more
recently, by Franco \& Strohmayer (1999) for a burst from 4U
1820--30 (see also Kuulkers et al. 2002a for a discussion). 
Highly significant discrete spectral components, which could
be modelled as an $\approx 6.4$ keV emission line and $\approx 8$ keV
absorption edge have recently been found in RXTE/PCA spectra during a
superburst from 4U 1820--30 (see Strohmayer \& Brown 2002). 
It seems likely that these features may result 
from reprocessing (disk reflection) and fluorescence in the accretion disk, 
or perhaps are formed in the burst-driven wind. It is not yet known if the 
features observed during PRE bursts are directly related to the features
observed during the 4U 1820--30 superburst.  Until their
identifications are more secure it will be difficult to infer neutron
star properties with them, however, these recent results have
established convincingly that some X-ray burst spectra do have
discrete lines. Observations with sufficient collecting area at higher
spectral resolution will likely provide the breakthrough needed for
reliable interpretation of these features.

\begin{figure*}
 \centering \vspace{0cm}
 \epsfig{file=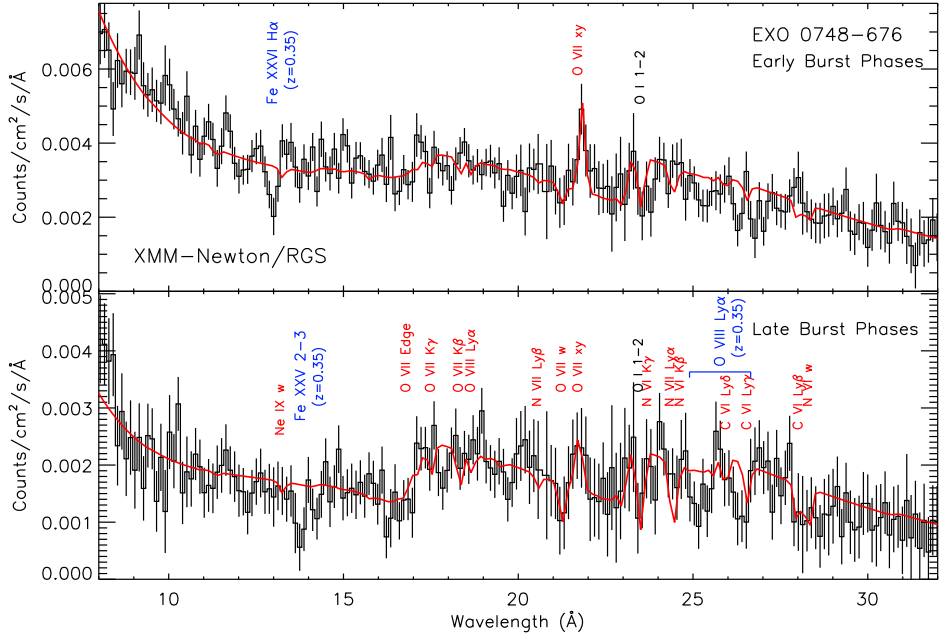,width=1.02\linewidth,angle=90} 
\caption{High resolution X-ray spectra of bursts from the LMXB EXO 
0748--676. The XMM-Newton Reflection Grating Spectrometer (RGS) data from
the early (top) and late (bottom) phases of the sum of 28 X-ray bursts are 
shown. The solid curve shows the best model of continuum and line features 
due to ionized gas in the vicinity of the neutron star. The unmodelled dips 
near 13 (top) and 14 (bottom) Angstroms are consistent with redshifted 
lines of Fe XXVI from the neutron star surface (after Cottam, Paerels \& 
Mendez 2002).}  
\label{figure5}
\end{figure*}

The high spectral resolution capabilities of Chandra and XMM/Newton
are providing new tools to study neutron star spectra, however, with
one recent exception (see Sanwal et al. 2002; Mereghetti et al. 2003),
the high resolution spectra of isolated neutron stars have been 
frustratingly devoid of line features (see for example, 
Walter \& Lattimer 2002; Burwitz et al. 2001; Drake et al. 2002; 
Pavlov et al. 2001). Indeed, bursters
may prove to be more promising targets for line searches because of
the mix of heavy elements constantly being provided by accretion. For
example, Cottam, Paerels \& Mendez (2002) have recently reported
evidence for redshifted absorption lines in XMM/Newton Reflection
Grating Spectrometer (RGS) data of bursts from EXO 0748--676. They co-added
data from 28 bursts in order to increase the sensitivity to narrow
lines. After modelling the continuum and line features thought to be
produced in the highly ionized gas surrounding the neutron star (see
Cottam et al.  2001), they found several features which were
unaccounted for by their best model. Figure 3.5, from Cottam, Paerels \&
Mendez (2002), shows their RGS spectra and best model. 
They interpreted the unmodelled features near 13 and 14 Angstroms
as redshifted absorption lines of the n = 2 -- 3 transitions of
hydrogen-like Fe (Fe XXVI) in the early (hotter) part of the bursts,
and the same transitions in the helium-like ion in the later (cooler)
portions of the bursts. The implied neutron star surface redshift in
each case was $z = 0.35$, and is consistent with modern neutron star
equations of state with reasonable masses (Lattimer \& Prakash
2001). It also implies that the neutron star surface is inside the
last stable circular orbit (Kluzniak \& Wagoner 1985). 

\subsection{Recurrence intervals and the $\dot M$ dependences} 
  
We previously discussed the one case where the observed type I bursts
matched theory well. However, most bursters are not so easy to
understand, and we discuss that more typical case here as it starts to
point to the possibility of non-spherically symmetric behavior.  In a
particular burning regime, we expect that the time between bursts
should decrease as $\dot M$ increases since it takes less time to
accumulate the critical amount of fuel at a higher $\dot M$.  Exactly
the {\it opposite} behavior was observed from many low accretion rate
($\dot M < 10^{-9} M_\odot \ {\rm yr^{-1}}$) NSs. A particularly good
example is 4U 1705--44, where the recurrence time increased by a factor
of $\approx 4$ when $\dot M$ increased by a factor of $\approx 2$
(Langmeier et al. 1987, Gottwald et al. 1989).  If the star is
accreting matter with $Z_{CNO}=0.01$ then these accretion rates are at
the boundary between unstable helium ignition in a hydrogen-rich
environment at high $\dot M$ and unstable pure helium ignition at
lower $\dot M$. The expected change in burst behavior as $\dot M$
increases would then be to more energetic and more frequent
bursts. This was not observed.

Other NSs showed similar behavior. Van Paradijs, Penninx \& Lewin
(1988) tabulated this effect for many bursters and concluded: {\it
``This suggests that continuous stable burning of a sizeable fraction
of the nuclear fuel is a general phenomenon on the surface of
accreting neutron stars. This fraction is apparently a gradually
increasing function of $\dot M$''}.  The following trends were found
as $\dot M$ increases:
\begin{itemize}
\item{The recurrence time increases from 2 -- 4 hours to 10 hours.}
\item{The bursts burn less of the accumulated fuel, with $\alpha$
increasing from $\approx 40$ to $>100$.}
\item{The duration of the bursts decreases from $\approx 30$ s to
$\sim 5$ s.}
\end{itemize}
\noindent 
More recent {\it RXTE} observations found the same trend in KS 1731-260 
(Muno et al. 2000), but not in 4U 1728--34 (Franco 2001; 
van Straaten et al. 2001). 

The low $\dot M$ bursts appear to be mixed H/He burning with a large H
mass fraction (i.e., energetic and of long duration from the
rp-process) whereas the high $\dot M$ bursts look more He dominated
with less H present (i.e., not so energetic, recurrence times long
enough to burn H while accumulating, and short duration due to the
lack of much energy release from the rp-process). The simplest
explanation is to say that the NS is near the transition from the low
$\dot M$ mixed burning regime (1 in \S 3.2.1) to the higher $\dot M$
pure helium burning (noted as 2 in \S 3.2.1). This would require that
the NSs are accreting at $\approx 10^{-10}M_\odot \ {\rm yr^{-1}}$ in
the lower $\dot M$ state and about a factor of 4 -- 5 higher in the high
$\dot M$ state. However, these estimates are very far away from that
observed.

Van Paradijs et al. (1988) used the ratio of the persistent flux to
the flux during Eddington limited PRE bursts as a measure of $\dot M$
in units of the Eddington accretion rate.  They showed that most
bursters accrete at rates $\dot M \approx (3-30)\times 10^{-10} \
M_\odot {\rm \ yr^{-1}}$, at least a factor of three (and typically
more) higher than the calculated rate where such a transition should
occur. Moreover, if the accretion rates were as low as needed, the
recurrence times for the mixed hydrogen/helium burning would be about
30 hours, rather than the observed 2 -- 4 hours.  Fujimoto et al. (1987)
discussed in some detail the challenges these observations present to
a spherically symmetric model, while Bildsten (1995) attempted to
resolve this by having much of the thermally unstable burning occur
via slow deflagration fronts that lead to very low frequency noise and
not Type I bursts (see also Yu et al. 1999). Whether this is possible
on rapidly rotating NSs is uncertain (Spitkovsky, Levin \& Ushomirsky
2002).

Much of this puzzle might be resolved by allowing the fresh material
to only cover a fraction of the star prior to igniting (Bildsten
2000).  There are strong observational clues that this may be
happening (e.g. Marshall 1982), as the other clear trend found by {\it
EXOSAT} was an increase in the apparent black-body radius ($R_{bb}$)
as $\dot M$ increased. This parameter is found by spectral fitting in
the decaying tail of the Type I bursts and, though susceptible to
absolute spectral corrections, can likely be trusted for relative
sizes (LVT).  In a similar vein, van der Klis et al. (1990) found that
the temperature of the burst at the moment when the flux was one-tenth
the Eddington limit decreased as $\dot M$ increased (hence a larger
area) for the Atoll source 4U 1636--53. In total, these observations
suggest the possibility that the covered area increases fast
enough with increasing $\dot M$ that the accretion rate per unit area
actually {\it decreases}. 

By interpreting the measured $R_{bb}$ as an indication of the
fraction of the star that is covered by freshly accreted fuel, the
quantity $\dot m=\dot M/4\pi R_{bb}^2$ can be calculated using 
$F_x=GM\dot M/4\pi d^2 R$ and gives $\dot m\approx
(F_xR/GM)(d/R_{bb})^2$.  Bildsten (2000) has argued that the radius
increase can offset the $\dot M$ increase in the context of data from
the burster EXO~0748--676 (Gottwald et al. 1986; see Figure 1 in
Bildsten 2000).

Where does the material arrive on the NS? 
We know that these NSs accrete from a disk formed in the Roche lobe
overflow of the stellar companion, however, there are still debates
about the ``final plunge'' onto the NS surface. Some advocate that a
magnetic field controls the final infall, while others prefer an
accretion disk boundary layer.
These arguments must also take into account the possible presence of 
an accretion gap between the inner disk edge
and the stellar surface (Kluzniak, Michelson \& Wagoner 1990, 
Kluzniak \& Wilson 1991), as the gravitational redshift measurement 
reported by Cottam et al. (2002) implies that the NS surface is inside 
the last stable orbit. If material is placed in the
equatorial belt, it is not clear that it will stay there very long. If
angular momentum was not an issue, the lighter accreted fuel (relative
to the ashes) would cover the whole star quickly.  However, on these
rapidly rotating NSs, the fresh matter added at the equator must lose
angular momentum to get to the pole. This competition (namely
understanding the spreading of a lighter fluid on a rotating star) has
been recently investigated by Inogamov \& Sunyaev (1999).

Another remaining conundrum is the burst behavior of the bright Z
sources (Sco X-1, Cyg X-2, GX 5-1, GX 17+2, GX 340+0, GX 349+2) which
are accreting at $3 \times 10^{-9} - 2\times 10^{-8}\ M_\odot \ {\rm
yr^{-1}} $. At the very upper end of this range, the burning could be
thermally stable, however, if unstable, they would clearly be in the
regime where the bursts are regularly spaced and of long duration due to
mixed hydrogen/helium burning (noted as 3 above and as exhibited by GS
1826--24).  However, this is not typically seen, rather, bursts are
rare and are clearly not responsible for burning all of the accreted
fuel. Though some of the recently studied bursts from GX 17+2
(Kuulkers et al. 2002b) do look as expected, not all are easy to
understand and we refer the interested reader to Kuulkers et
al. (2002b) for an excellent summary discussion of the current
observational situation at high $\dot M$'s.

Most of the Type I bursts we have discussed occur when the neutron
star is accreting at a rate in excess of $10^{-10} M_\odot \ {\rm
yr^{-1}}$.  Recently, however, the BeppoSAX WFCs have discovered a
number of new thermonuclear burst sources from which no persistent
X-ray flux could be detected down to the $10^{-10}$ erg cm$^{-2}$
s$^{-1}$ flux limit of the WFCs (see for example, Cornelisse et
al. 2002a; Cocchi et al. 2002; in 't Zand et al.  2002). If these
objects are closer than 10 kpc then their accretion rate when bursting
was less than $10^{-10} M_\odot \ {\rm yr^{-1}}$ (or $L<10^{36} \ {\rm
erg \ s^{-1}}$).

At other epochs, some of these sources were observed
by ROSAT at luminosities $\approx 10^{35}$ erg s$^{-1}$, whereas later
Chandra observations of several of them revealed persistent
luminosities (or upper limits) in the $10^{32-33}$ erg s$^{-1}$ range
(Cornelisse et al. 2002b), consistent with flux levels seen from
neutron star X-ray transients in quiescence (see for example Rutledge
et al. 2001). Cocchi et al. (2001) and Cornelisse et al. (2002b) have
suggested that these bursters may be either a new class of low
persistent emission bursters (in which case there is a huge underlying
population) or weak transients with time-averaged accretion rates
$<10^{-11}M_\odot \ {\rm yr^{-1}}$ (like SAX J1808.4--3658) that just
happen to be further away.

Regardless of whether these systems are constantly accreting at such a
low rate or are weak transients in outburst, what is important for
this discussion is that the bursts occur when the accretion rate is
clearly less than $10^{-10}M_\odot \ {\rm yr^{-1}}$.  Presuming
complete covering of the neutron star, the hydrogen burning would be
unstable at these accretion rates (regime 1 in \S 3.2.1 )
and the unstable ignition should lead to long (100 -- 1000 s),
infrequent bursts (recurrence times of $\approx$ weeks).  Some of the
recently discovered sources appear to be in this regime (Kaptein et
al. 2000; in 't Zand et al.  2002), however, it is not known if all
these ``burst-only'' sources can be explained this way. In fact, it
seems unlikely, since some of the observed bursts are short and
therefore inconsistent with the standard low accretion rate theory. It
is conceivable that our understanding of these sources may require
additional, important physics, such as element diffusion, which is
not important for higher accretion rates but becomes more relevant
at these rates (Wallace et al. 1982, Bildsten, Salpeter \& Wasserman 1993). 

\section{Millisecond variability during X-ray bursts}

Early theoretical studies noted the likely importance of spreading of
the thermonuclear burning front around the neutron star surface (see
for example, Joss 1978). Because nuclear fuel is burned in a time much
shorter than it takes to accrete a critical pile it is unlikely
that ignition conditions will be achieved over the entire surface
simultaneously (Shara 1982).  It appears more likely that burning is
initiated locally and then spreads laterally, eventually engulfing all
fuel-loaded parts of the neutron star.  For conditions most prevalent
in burst sources, the front may spread via convective
deflagration, at lateral speeds of up to $\approx 5 \times 10^6$ cm
s$^{-1}$ (see Fryxell \& Woosley 1982; Hanawa \& Fujimoto 1984;
Nozakura, Ikeuchi \& Fujimoto 1984; Bildsten 1995). Such speeds can
account for the sub-second rise times of some bursts, but the time
required for burning to engulf the entire star is still long compared
to the spin periods of accreting LMXB neutron stars
(milliseconds). Moreover, if the burning front is not strongly
convective, then a patchy distribution of nuclear fuel is possible
(see Bildsten 1995). These considerations suggest that during bursts
the rotation of the neutron star can modulate the inhomogeneous or
localized burning regions, perhaps allowing for direct observation of
the spin of the neutron star.

Mason et al. (1980) reported a 36.4 Hz pulsation in an optical burst
from 4U 1254--690, and Sadeh (1982) claimed detection of a 12 ms
modulation in a burst from 4U 1728--34 with HEAO-1, but these periods
were never confirmed by subsequent observations. Murakami et al.
(1987) reported 1.5 Hz oscillations during the PRE phase of a burst
from 4U 1608--52. These oscillations might conceivably be caused by
oscillations of the photospheric radius at constant luminosity
(Lapidus et al. 1994). Schoelkopf \& Kelley (1991) reported detection
at the $\approx 4\sigma$ level of a 7.6 Hz oscillation in Einstein
Monitor Proportional Counter data during a burst from Aql X-1. They
suggested that rotation of the neutron star with a non-uniform surface
brightness might be responsible, but the signal has not been seen in
other bursts. Jongert \& van der Klis (1996) searched for high
frequency variability in bursts observed with the EXOSAT
observatory. They averaged power spectra from multiple bursts from the
same source to increase sensitivity, but also found no significant
periods. They placed upper limits on average modulation amplitudes in
the $\sim$ 100 Hz range of between 5 - 10 \%.

\subsection{Overview of burst oscillations}

An exciting development in the past decade has been the discovery of
high frequency (300 -- 600 Hz) X-ray brightness oscillations during
bursts.  These modulations are now commonly called ``burst
oscillations.''  They were first discovered with the PCA onboard RXTE
in bursts from the LMXB 4U 1728--34 (Strohmayer et al. 1996). As of
this writing, burst oscillation detections have been claimed for an
additional ten sources, one of which is the 401 Hz accreting
millisecond pulsar SAX J1808.4--3658 (in 't Zand et al. 2001; Wijnands et
al. 2002), whose spin period is precisely known. Table 3.1 provides a
catalog of the known burst oscillation sources, summarizes some of
their salient properties, and provides references to the relevant
literature.

\begin{figure*}
 \centering \vspace{1cm}
 \epsfig{file=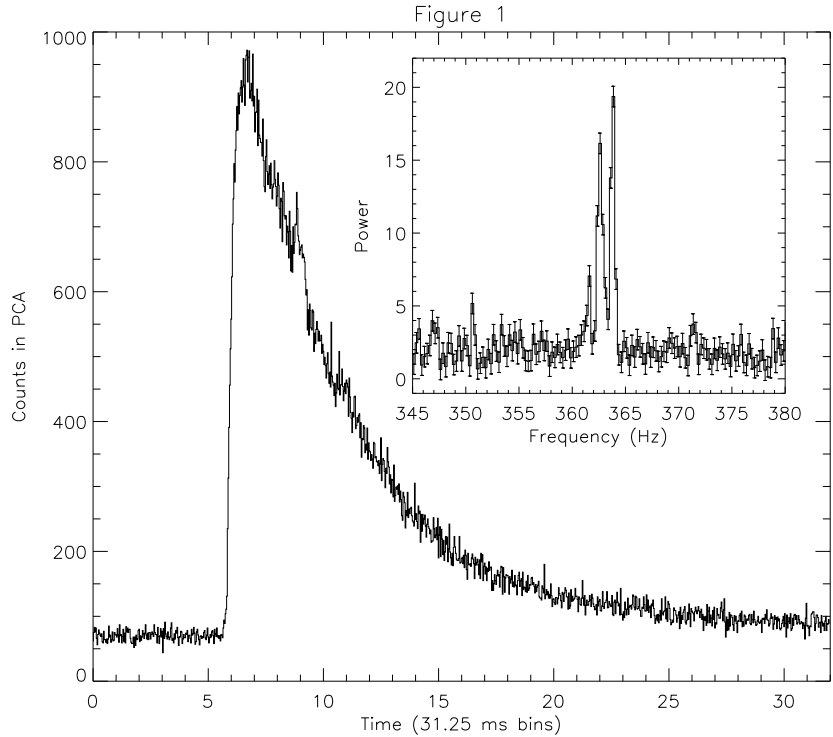,width=0.85\linewidth} 
\caption{An X-ray burst from 4U 1728--34 observed with the PCA onboard
RXTE.  The main panel shows the X-ray counts observed by the PCA in
(1/32) s bins. The inset panel shows the power spectrum in the
vicinity of 363 Hz (after Strohmayer et al. 1996).}  
\label{figure6}
\end{figure*}

Burst oscillations were first discovered in Fourier power spectra
computed from high time resolution lightcurves of entire
bursts. Figure 3.6 shows an example of a Fourier power density
spectrum of a burst from 4U 1728--34 with a burst oscillation at 363 Hz
(after Strohmayer et al. 1996). Although many detailed questions
remain, there is now little doubt that spin modulation of the X-ray
burst flux is the basic mechanism responsible for these oscillations.
The discovery of burst oscillations was closely linked with the
discovery of millisecond variability (kHz QPOs) in the persistent
X-ray flux from accreting neutron stars. Such oscillations, and some
of their inferred connections to burst oscillations, are reviewed in
Chapter 2 by van der Klis. Therefore, we will only review the
observational properties and current theoretical understanding of
burst oscillations and will not discuss in any great detail the
possible connections between the two phenomena.  

\begin{table*}[tbp]
\begin{center}
 \caption{Burst oscillation sources and properties}
  \begin{tabular}{ccccc}
   \hline \hline
Source & Frequency (Hz) & $\Delta\nu_{qpo}^2$ & $P_{orb} ({\rm hr})^3$ & 
References$^1$
\\
\hline
4U 1728--34 & 363 & 280 -- 363 & ? & 1,2,3,4,5,13,14 \\
4U 1636--53 & 581 & 250 -- 320 & 3.8 & 6,7,25,26,29 \\
KS 1731--260 & 524 & 260 & ? & 10,11,12 \\
Galactic Center & 589 & ? & ? & 15 \\
Aql X-1 & 549 & ? & 19.0 & 16,17 \\
4U 1702--429 & 330 & 315 -- 344 & ? & 4,9 \\
MXB 1658--298 & 567 & ? & 7.1 & 18, 27 \\
4U 1916--053 & 270 & 290 -- 348 & 0.83 & 19,20 \\
4U 1608--52 & 619 & 225 -- 325 & ? & 8, 21 \\
SAX J1808.4--3658 & 401 & ? & 2.0 & 22,23,28 \\
SAX J1750.8--2980 & 601 & ? & ? & 24 \\
\hline \hline
\end{tabular}
\end{center}
\begin{minipage}{0.99\linewidth}
$^{1}$ {\rm (1) Strohmayer et al. (1996); (2) Strohmayer,
Zhang, \& Swank (1997); (3) Mendez \& van der Klis (1999); (4) Strohmayer 
\& Markwardt (1999); (5) Strohmayer et al. (1998b); (6) Strohmayer et al. 
(1998a); (7) Miller (1999); (8) Mendez et al. (1998); (9) Markwardt, 
Strohmayer \& Swank (1999) (10) Smith, Morgan, \& Bradt (1997); 
(11) Wijnands \& van der Klis (1997); (12) Muno et al. (2000); (13) van 
Straaten et al. (2000); (14) Franco (2000); (15) Strohmayer et al (1997); 
(16) Zhang et al. (1998); (17) Ford (1999); (18) Wijnands, Strohmayer \& 
Franco (2001); (19) Boirin et al. (2000); (20) Galloway et al. (2001); (21) 
Chakrabarty (2000); (22) in 't Zand et al. (2001); (23) Ford (2000); (24) 
Kaaret et al (2002); (25) Giles et al. (2002); (26) Strohmayer \& Markwardt 
(2002); (27) Wijnands et al. (2002); (28) Wijnands et al. (2002); (29) 
Jonker, Mendez \& van der Klis (2002).}

$^2$ The frequency separation between pairs of kHz QPO in Hz, if known.

$^3$ Orbital period of the system, in hours, if known.

\end{minipage}
\label{table1}
\end{table*}

\subsection{Oscillations during the burst rise}

In some bursts oscillations are detected during the rising portion of
the burst time profile. Indeed, detections can be made within a few
tenths of a second after a significant rise in the X-ray flux is seen
(see Strohmayer et al. 1988a; Miller 1999). Such oscillations can have
very large amplitudes. Strohmayer, Zhang \& Swank (1997) found that
some bursts from 4U 1728--34 show oscillation amplitudes as large as 43\% 
within 0.1 s of the onset of the burst. Strohmayer et al. (1998a)
found bursts from 4U 1636--53 with modulations at onset of $\approx 75
\%$ (see Figure 3.7). These early studies also found that the
modulation amplitude is anti-correlated with the X-ray intensity
during the rise. That is, the amplitude drops as the flux increases
toward maximum. This behavior is consistent with simple expectations
for spin modulation of an initially localized X-ray ``hot spot'' which
expands in $\sim 1$ s to engulf the neutron star. That is, the
amplitude is largest when the hot spot is smallest, at onset, and then
decreases as the spot engulfs the star and the flux increases (see Nath, 
Strohmayer \& Swank 2002).  Time resolved X-ray spectroscopy during 
bursts provides additional evidence
for localized X-ray emission near burst onset. For example,
Strohmayer, Zhang \& Swank (1997) tracked the evolution of the
bolometric flux and blackbody temperature during bursts from 4U
1728--34. For a spherical blackbody source, the quantity
$F_{bol}^{1/4} / kT_{bb}$ is proportional to the square root of the
emitting area (see also the discussions in \S 3.3.1 and 3.3.4).  
They found that this inferred surface area was
smallest at onset and then increased as the burst flux increased
during the rise. The inferred area then stayed approximately constant
as the flux declined (see Figure 3.8).

\begin{figure*}
 \centering 
 \centerline{\epsfig{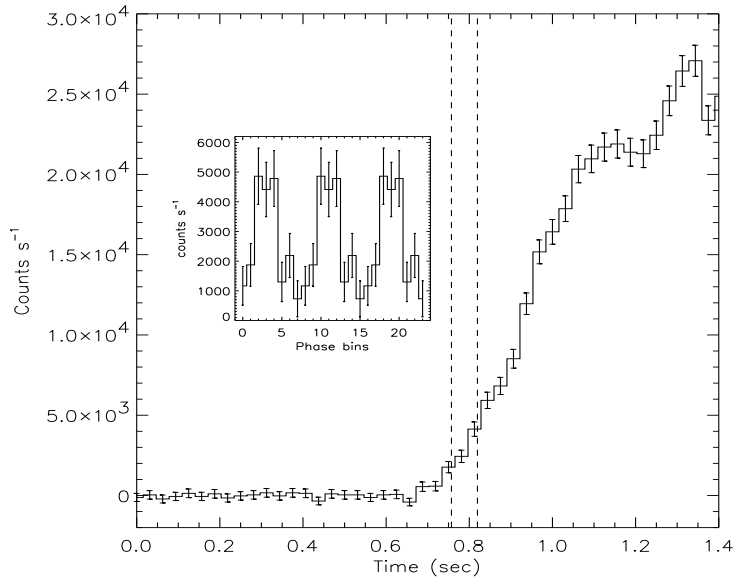}}
\caption{X-ray timing evidence indicating a spreading hot spot at the onset 
of thermonuclear bursts. The main panel shows a burst from 4U 1636--53 with 
large amplitude, 581 Hz oscillations on the rising edge of the profile. 
The inset shows the pulse profile during the interval marked by the vertical 
dashed lines.  The pulse profile is repeated $3 \times$ for clarity. 
Note the large amplitude of the oscillation. (after Strohmayer et al. 1998a).}
\label{figure7}
\end{figure*}

\begin{figure*}
 \centering 
 \centerline{\epsfig{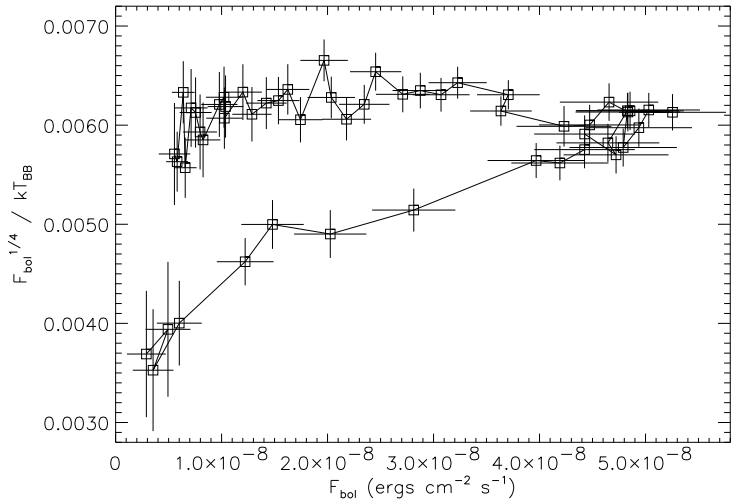}}
\caption{X-ray spectral evidence indicating a
spreading hot spot at the onset of thermonuclear bursts.   
The plot shows the time evolution of the inferred surface area in a burst 
from 4U 1728--34. The evolution is from lower left to upper right, and then
horizontally to the left (after Strohmayer, Zhang \& Swank 1997). }
\label{figure8}
\end{figure*}

\subsection{Oscillations in the decay phase}

Oscillations are also observed at late times in the cooling or decay
phase of bursts. In many cases oscillations are seen after
``touchdown'' in PRE bursts (see for example, Smith, Morgan \& Bradt
1996; Muno et al. 2000), however, bursts without PRE can also show
oscillations which persist during the cooling phase (Franco 2001; van
Straaten 2001). Typically the amplitude of oscillations seen in the
cooling phase is lower than observed during burst rise, but amplitudes
as large as 15 \% (rms) have been observed. Some bursts show
oscillations both on the rise and in the cooling phase (see van
Straaten 2001; Franco 2001), however, oscillations are not detected
during PRE phases, only before and after the PRE episode (see Muno et
al. 2002). This provides additional evidence that the modulations are
associated with processes on the neutron star surface. In bursts where
pulse trains are detected during the rise and decay they can sometimes
be observed for as long as $\approx$ 12 seconds. The presence of strong 
oscillations after the nuclear burning has presumably 
engulfed the entire neutron star is difficult to account for in the simplest 
expanding hot spot scenarios. This has motivated a number of recent 
theoretical ideas to explain the persistence of burst oscillations. We will 
discuss some of these ideas shortly. 

\subsection{Frequency evolution of burst oscillations}

The oscillation frequency during a burst is typically not constant.
Most commonly the frequency increases as the burst progesses, that is,
the evolution can be characterized as a ``chirp.'' Most bursts show
spin-up toward some limiting, or asymptotic frequency, however, there
are exceptions to this rule of thumb. For example, Strohmayer (1999)
and Miller (2000) identified a burst from 4U 1636--53 with spin down in
the cooling tail.  This burst also showed an unusually long thermal
tail which may have been related to a ``reheating'' episode having
some connection to the spin down. Muno et al. (2000) found spin down
in a burst from KS 1731--260, however, they found no evidence for
unusual flux enhancements or spectral variations during the
episode. Spin downs are apparently rare. In a recent study, Muno et
al. (2002) found some evidence for them in only 3 of 68 bursts
examined.  The observed frequency drifts are generally $< 1\%$ of the
mean frequency, and bursts which have detectable pulsations during the
rising phase show the largest frequency shifts. This indicates that
the process responsible for the frequency evolution begins with the
start of a burst, and not when oscillations are first detected within
a burst (Muno et al. 2002).

Strohmayer \& Markwardt (1999) studied the frequency evolution in
bursts from 4U 1702--429 and 4U 1728--34. They found the frequency in
these bursts could be modelled as a smooth exponential recovery of the
form, $\nu (t) = \nu_0 ( 1 - \delta_{\nu} e^{-t/\tau} )$, where
$\nu_0$, $\delta_{\nu}$, and $\tau$ are the asymptotic frequency, the
fractional frequency drift, and the recovery timescale, respectively.
With this form they were able to recover coherent signals, with
coherence values, $Q \equiv \nu_0 / \Delta\nu_0 > 4,500$ in some
bursts. Figure 3.9 shows an example of a burst from 4U 1702--429 with
exponential frequency evolution.  These results support the existence
of a reference frame on the neutron star, perhaps the nuclear burning
layer, in which the oscillations are coherent or nearly so. This
frame, however, cannot be rigidly connected to the bulk of the neutron
star, because the torque required to change the spin frequency of the
star by $\approx 1 \%$ in only 10 seconds is unphysically large. This
implies the existence of shearing in the surface layers of the neutron
star. In the exponential model, the total amount of phase shearing is
simply $\phi_{shear} = \nu_0\delta_{\nu} ( 1 - e^{-T/\tau} )$, where
$T$ is the length of the pulse train.  For typical bursts this value
ranges from about 4 -- 8, suggesting that the burning layer ``slips"
this many revolutions over the underlying neutron star during the
duration of the pulsations. The amount of phase shearing has
implications for the surface magnetic field strength, as a
sufficiently strong field will enforce co-rotation (Cumming \&
Bildsten 2000).

Muno et al. (2000) carried out a detailed study of the frequency
evolution in burst oscillations from KS 1731--260. They used pulsar
timing techniques to compute phase-connected timing solutions in order
to study the functional form of the frequency evolution. They found
that the phase evolution can usually be modelled as a polynomial, and
that the exponential relaxation model was adequate for many bursts.
More recently, Muno and colleagues (2002) have explored the frequency
evolution in a larger sample of bursts (68) from several sources. They
confirm many of their earlier findings, however, they do find a subset
of bursts for which simple 2nd and 3rd order polynomials are
insufficient to explain the phase evolution. In these bursts there is
evidence for phase jitter on timescales of seconds. Figure 3.10 shows
several examples of burst oscillations with complex phase
evolution.  They suggest this may indicate the presence of some
instability in the mechanism which generates the oscillation. One idea
is that two signals, or modes, with nearly equal frequency may be
simultaneously present (see also Miller 1999). Another possibility is
that phase jumps occur on relatively short timescales (see Strohmayer
2001).  These behaviors, though infrequent, may provide important
clues to the physical mechanism which produces the observed
modulations, particularly in the cooling phase.

Strohmayer et. al (1997) argued that the observed frequency evolution
results from angular momentum conservation of the thermonuclear shell.
Burst-induced heating expands the shell, increasing its rotational
moment of inertia and slowing its spin rate. Near burst onset the
shell is thickest and thus the observed frequency lowest. The shell
spins back up as it cools and recouples to the underlying neutron
star. Calculations indicate that the $\sim 10$ m thick pre-burst shell
can expand to $\sim 30$ m during the flash (see Joss 1978; Bildsten
1995; Cumming \& Bildsten 2000), which gives a frequency shift of
$\approx 2 \ \nu_{spin} (20 \ {\rm m}/ R)$, where $\nu_{spin}$ and $R$
are the stellar spin frequency and radius, respectively. For typical
burst oscillation frequencies this gives a shift of $\sim 2$ Hz,
similar to that observed. However, Galloway et al. (2000) reported a
3.5 Hz frequency shift in a burst from 4U 1916--053 with 272 Hz
oscillations. They suggested that such a large change, $\sim 1.3 \%$
might be inconsistent with expansion of the thermonuclear burning
layer because of the magnitude of the implied height change of $\sim
80$ m. Wijnands, Strohmayer \& Franco (2001) found a $\sim 5$ Hz
frequency shift in a burst from 4U 1658--298 with a 567 Hz oscillation,
which may also be uncomfortably large given current estimates of the
expansion of the burning layers (Cumming et al. 2002).

\begin{figure*}
 \centering 
 \epsfig{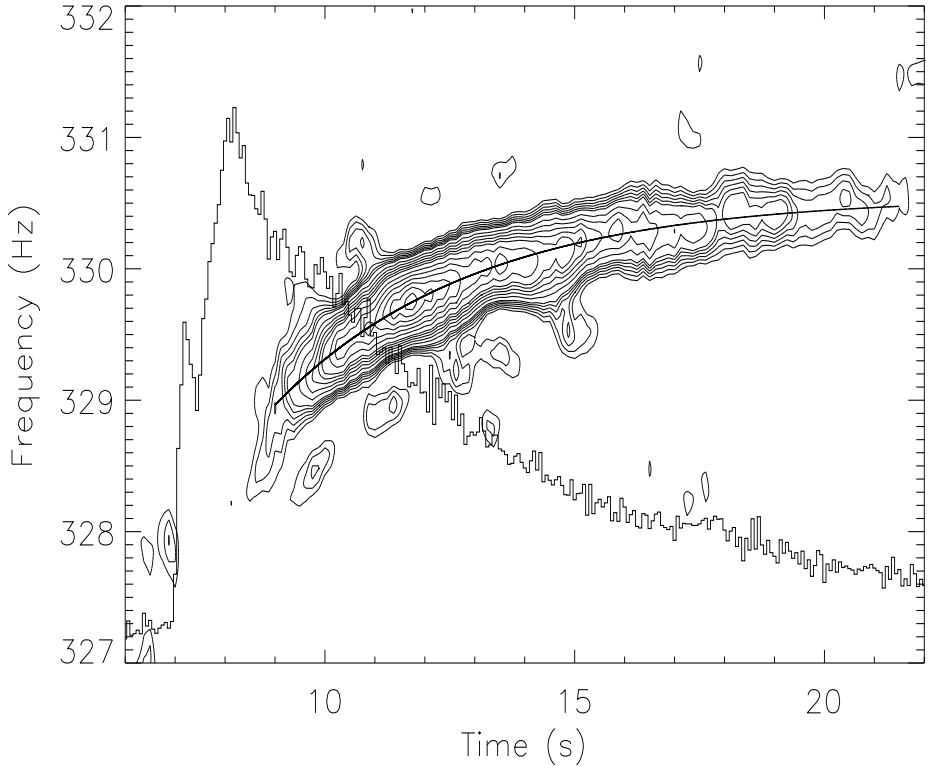} 
 \caption{An X-ray burst from 4U 1702--429 observed with the PCA onboard RXTE. 
Shown are contours of constant power spectral density as a function of 
frequency and time. The solid curve shows the best fitting exponential model. 
The burst time profile is also shown (after Strohmayer \& Markwardt 1999).}  
\label{figure9}
\end{figure*}

\subsection{Long term stability of burst oscillation frequencies}

Studies of the frequency stability of burst oscillations over years
provides constraints on the mechanism which sets the
frequency. Strohmayer et al. (1998b) carried out one of the first
studies of long term stability. They compared the asymptotic burst
oscillation frequencies in bursts from 4U 1728--34 and 4U 1636--53.  For
bursts from 4U 1728--34 spanning a 1.6 year epoch they found that the
asymptotic pulsation period was the same to better than 1
$\mu$sec. This suggests a timescale to change the frequency longer
than 23,000 years, and supports a mechanism like rotation with a high
degree of intrinsic stability.
 
The accretion-induced rate of change of the neutron star spin
frequency in a LMXB has a characteristic value of $1.8 \times 10^{-6}$
Hz yr$^{-1}$ for a canonical neutron star and typical mass accretion
rates for X-ray burst sources. Over a year the accretion induced shift
is much smaller than the apparent spin frequency changes which would
be caused by the projected orbital motion of the neutron star. This
led to the suggestion that, if burst oscillation frequencies were
intrinsicaly stable enough, one might be able to extract the projected
orbital velocity of the neutron star from a sample of bursts observed
at different orbital phases.  Recent studies, however, have confirmed
that it will be much more difficult to extract neutron star velocity
information from burst oscillation frequencies than initially hoped.
For example, Giles et al. (2002) studied the burst oscillation
frequencies of 26 bursts from the LMXB 4U 1636--53. The highest
observed oscillation frequency of all bursts in their study are stable
at the level of $2 \times 10^{-3}$, but are not correlated with
orbital phase as expected for binary modulation.

\begin{figure*}
 \centering
 \epsfig{file=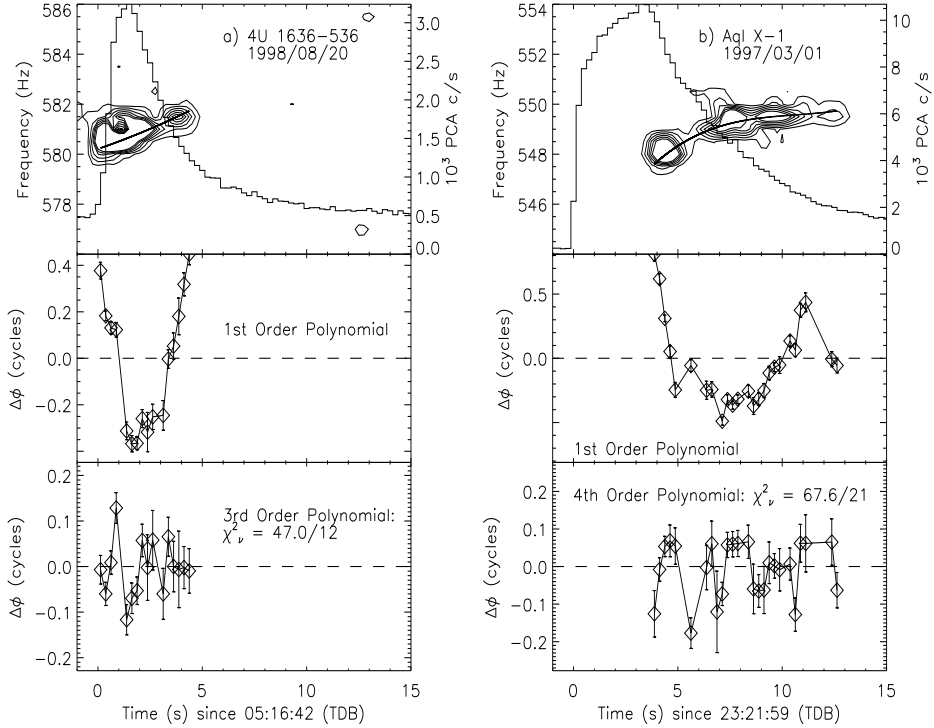,height=4.0in,width=0.99\linewidth}
 \caption{Examples of two burst oscillations which show complex phase 
evolution. In both cases higher order polynomial phase evolution functions
(3rd (left) and 4th (right) order) still leave significant residuals (after
Muno et al. 2002).}
\label{figure10}
\end{figure*}

In a related study Muno et al. (2002) examined the asymptotic burst
oscillation frequencies of bursts from 8 different sources, including
4U 1636--53. They quantify the dispersion in asymptotic frequencies
using the standard deviation, $\sigma_{\nu}$, of the observed
frequencies and find that $\sigma_{\nu} / < \nu_{max} >$ is typically
$< 1 \times 10^{-3}$. Figure 3.11 summarizes the distributions of
asymptotic burst oscillation frequencies found in the Muno et
al. (2002) study. These results indicate that the asymptotic burst
oscillation frequencies are quite stable, but that there is more
variation than can easily be accounted for by binary Doppler
modulations alone. This requires that models for the burst
oscillations must be able to account for an intrinsic fractional
frequency variation perhaps as large as $\approx 5 \times 10^{-4}$.

One physical effect that approaches this magnitude is the
change in the thickness of the burning layer from before the burst to
after the burst due to the change in mean molecular weight (Cumming \&
Bildsten 2000, see their figure 12). A complete burn of hydrogen via the 
rp-process can cause a fractional change in the rotation rate of 
$\approx 5\times 10^{-4}$ (Cumming et al. 2002), so that varying levels of 
burning can lead to differing asymptotic frequencies at this level. 

\begin{figure*}
 \centering
 \epsfig{file=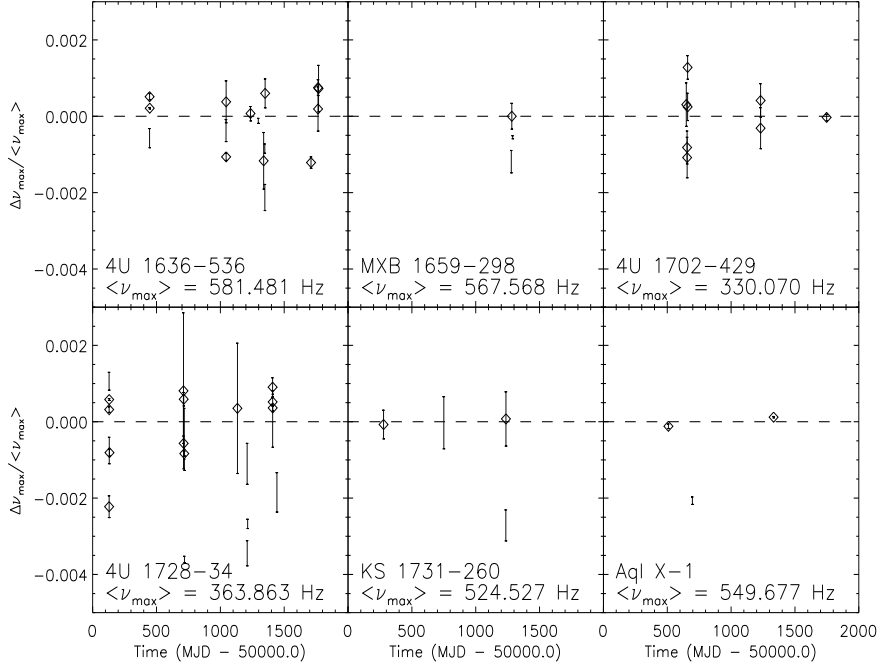,height=4.0in,width=0.99\linewidth}
 \caption{The fractional deviations from the mean of maximum burst oscillation 
frequencies from 6 different burst sources. The data are shown as a function 
of time. The standard deviation, $\sigma_{\nu} / <\nu_{max}>$, is typically 
$7 - 10 \times 10^{-4}$ (after Muno et al. 2002).}
 \label{figure11}
\end{figure*}

\subsection{Burst oscillations and the mass accretion rate}

Not all bursts from a given source have detectable oscillations.
Recent work indicates that the mass accretion rate, $\dot M$, onto the
neutron star has a strong influence on the strength, and therefore the
detectability of burst oscillations. This is perhaps not too
surprising, since $\dot M$ is known to influence other burst
properties as well, and, as outlined earlier, is an important
ingredient in any description of the nuclear burning physics. It is
not unreasonable then to expect that the properties of burst
oscillations might also depend importantly on $\dot M$. 

Muno et al. (2000) carried out the first systematic study of burst
oscillation properties and source spectral state.  They studied bursts
from the LMXB and atoll source KS 1731--260. The 524 Hz burst
oscillations in this source were discovered by Smith, Morgan \& Bradt
(1997). Muno et al. (2000) studied both the spectral and timing
properties of bursts as a function of the position of the source in an
X-ray color-color diagram (CD), and found that burst properties,
including the presence or absence of burst oscillations, were strongly
segregated in the CD. In particular, they found that only bursts which
occurred when the source was located on the ``banana branch'' of the
atoll pattern (i.e. at high inferred $\dot M$) produced detectable
oscillations. These bursts also showed PRE, had the highest peak
fluxes and had characteristically short durations (so called ``fast''
bursts). As discussed earlier these latter characteristics are an
indication that helium is the primary fuel in such bursts (see \S
3.2.2).

\begin{figure*}
 \centering
 \epsfig{file=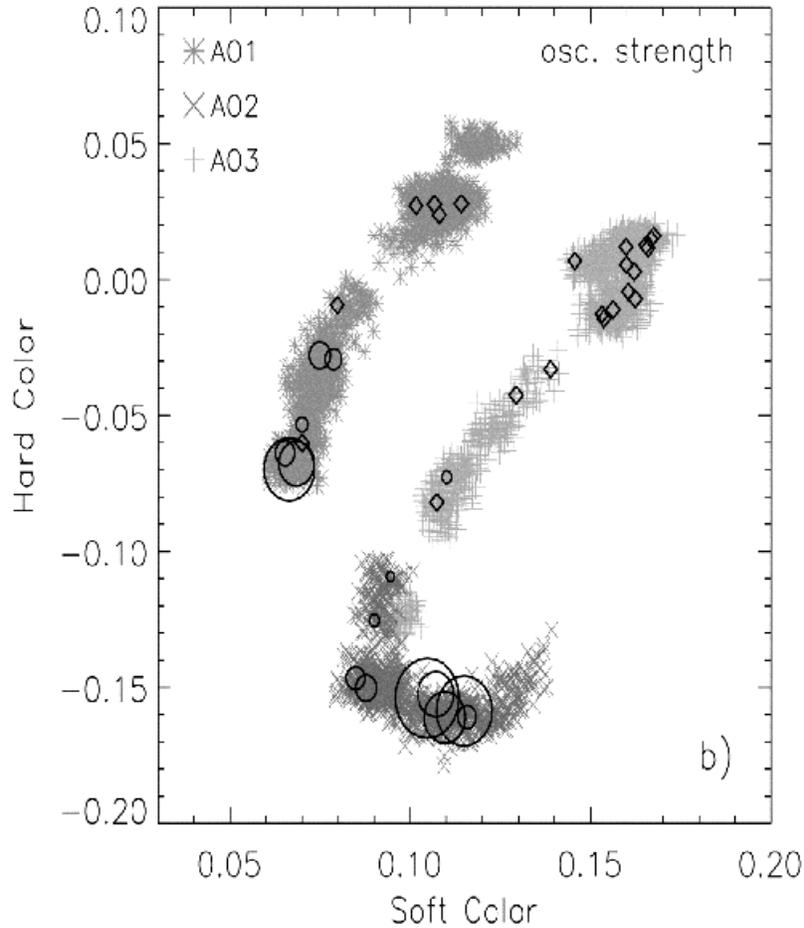,height=5.0in,width=0.85\linewidth}
 \caption{X-ray color-color diagram for 4U 1728--34 showing a typical atoll 
track identifying, counterclockwise from upper right, the extreme island
state (EIS), the island state (IS), the lower banana branch (LB), and the
upper banana branch (UB). The locations of bursts both with (circles) and 
without (diamonds) oscillations are shown. For bursts with oscillations, 
the size of the circle is proportional to the integrated strength of the 
oscillations (after Franco 2001).}
 \label{figure12}
\end{figure*}

Subsequent studies of bursts from 4U 1728--34 by van Straaten et
al. (2001) and Franco (2001) confirmed some of the main results found
by Muno et al. (2000), but new findings also complicated the
picture. For example, both of these studies found a similar dependence
of burst oscillations on source position in the CD. That is, bursts
with oscillations are restricted to the banana branches at higher
inferred $\dot M$. However, the properties of so called ``fast'' and
``slow'' bursts were not as nicely segregated in 4U 1728--34 as for KS
1731--260. For example, the bursts at low inferred $\dot M$ (in the so
called island state of the CD) from 4U 1728--34 all showed PRE whereas
the corresponding bursts from KS 1731--260 do not. Moreover, some of
the ``fast'' bursts from 4U 1728--34 which show strong burst
oscillations show no evidence for PRE. Franco (2001) further
characterized the bursts by computing a measure of integrated
oscillation strength through the bursts. This integrated oscillation
strength increased with inferred $\dot M$ on the CD (see Figure 3.12). 
A relationship between position in the CD and where during a burst 
oscillations are detected was also suggested. Bursts with oscillations 
detected only during the rising phase were found at the highest inferred 
$\dot M$, while bursts with oscillations only in the cooling (decay) phase 
were found at lower $\dot M$ (but still on the banana branch). 

\subsection{Burst oscillations and photospheric radius expansion}

Soon after the discovery of the first burst oscillation sources a
connection between the appearance of oscillations in bursts and PRE
was apparent.  Smith, Morgan \& Bradt (1997) found that oscillations
in a PRE burst from KS 1731--260 were first detected after photospheric
touchdown. Strohmayer et al. (1997) found similar results for bursts
with 589 Hz oscillations from a burster near the Galactic center. Muno
et al. (2000) have explored the connection between PRE and
oscillations in a large sample of bursts from several sources.  They
segregated burst oscillation sources into two classes, those with
burst oscillation frequencies closer to 300 Hz (so called ``slow''
oscillators), and those with frequencies close to 600 Hz (the ``fast''
group), and found that those bursts with fast oscillations were almost
always observed in bursts with PRE episodes, while bursts with slow
oscillations only showed PRE about half the time.  Muno et al. (2000)
suggest that this distinction between the fast and slow sources cannot
be an observational selection effect. They argue that the distinction
could result if burst properties vary differently with $\dot M$ in the
fast and slow sources.  Exactly how this difference comes about and
how it is related to the oscillation (or spin) is not yet understood.

\subsection{Harmonics, subharmonics and sidebands}

The pulse shapes of burst oscillations are highly
sinusoidal. Strohmayer \& Markwardt (1999) found no evidence for
harmonics after coherently summing oscillation signals from 4U
1728--34 and 4U 1702--429. They placed limits on the ratio of pulsed
amplitude at the fundamental to that at the first harmonic of $\approx
24$ and 15 in these sources, respectively. Muno et al. (2000), found
only weak ($2\sigma$) evidence of a signal at the first harmonic of
the 524 Hz frequency in bursts from KS 1731--260.  Recently, Muno,
$\ddot {\rm O}$zel \& Chakrabarty (2002) have explored the amplitude
evolution and harmonic content of cooling phase burst oscillations in
eight different sources. They find mean amplitudes (rms) of about $5
\%$, and during a typical burst the amplitude can vary in a manner
which is uncorrelated with the burst flux. They did not detect any
harmonic, nor subharmonic signals, and placed upper limits on the
fractional amplitudes at integer and half-integer harmonics of less
than 5 \% and 10 \% of the amplitude of the strongest observed signal,
respectively. Comparison of these results to theoretical models with
one or two circular, antipodal hot spots suggests that if a single
spot is present it must lie near the rotational pole or cover a
substantial fraction $(\approx 1/2)$ of the neutron star in order to
be consistent with the limits on harmonic signals.  If antipodal spots
are present, then the implications are that the spots must lie close
to the rotational equator.

Chakrabarty (2002) has found evidence for sidebands separated by 30 --
50 Hz from the burst oscillation frequency in a few sources.  These
sidebands do not yet fit cleanly within the context of current models,
however, Spitkovsky et al. (2002) have argued that zonal flows
associated with burst heating and rapid spin of the neutron star may
produce a modulation pattern as well as sidebands (see \S 3.4.11).

In many LMXBs a pair of kHz QPOs are observed (see the review by van
der Klis in Chapter 2). In six sources the frequency difference
between the kHz QPO is close to the observed burst oscillation
frequency or one half of the frequency. This closeness of the burst
oscillation frequency, and the frequency difference of the kHz QPOs
has motivated the ``beat frequency'' models of kHz QPOs (see
Strohmayer et al. 1996; Miller, Lamb \& Psaltis 1998; Lamb \& Miller
2001).  In three sources the QPO difference frequency is close to half
the burst oscillation frequency. This might be possible if a pair of
antipodal hot spots on the neutron star produces the burst
oscillation. Based on these considerations, Miller (1999) searched for
a subharmonic of the 581 Hz burst oscillation in 4U 1636--53, and
claimed detection of a 290 Hz signal at a $4\times 10^{-5}$
significance level. Extensive efforts to detect such a signal again
have, however, been unsuccessful (see Strohmayer 2001 for a
discussion).

\subsection{Burst oscillations as probes of neutron stars}

Modelling of burst oscillations holds great promise as a new tool for
probing the structure of neutron stars and their environs. The
emission and propagation of photons from the surfaces of rapidly
rotating neutron stars are strongly influenced by General relativistic
effects.  Gravitational light deflection suppresses the modulation
amplitude and reduces the harmonic content of pulses produced by
rotational modulation of a hot spot on a rotating neutron star (see
Pechenik, Ftaclas \& Cohen 1983). The strength of
light deflection is a function of the compactness, $m/r \equiv GM/c^2
R$. More compact stars produce greater deflections and therefore
weaker spin modulations (see Strohmayer et al. 1998a; Miller \& Lamb
1999). Relativistic motion of the hot spot creates asymmetry and
sharpening of the pulse profile, increasing the harmonic content (Chen
\& Shaham; Miller \& Lamb 1999; Braje, Romani \& Rauch 2000; Weinberg,
Miller \& Lamb 2001). Such motion also introduces a pulse phase
dependent Doppler shift in the X-ray spectrum.  The magnitude of these
effects are directly proportional to the surface velocity, which is a
function of the unknown stellar radius and the known spin
frequency. Ford (1999) has analysed data during a burst from Aql X--1
and finds that the softer photons lag higher energy photons in a
manner which is qualitatively similar to that expected from a rotating
hot spot. Fox (2001), however, found that the sense of the phase lags
in this burst switched, with hard lags preceding the soft lags found
by Ford et al. (1999). This suggests that photon scattering may play
an important role in addition to Doppler beaming.

Several studies have been undertaken to constrain neutron star
properties based on modelling of burst oscillations. Miller \& Lamb
(1998) have investigated the amplitude of rotational modulation
pulsations as well as harmonic content assuming emission from a
point-like hot spot. They also showed that knowledge of the angular
and spectral dependence of the emissivity from the neutron star
surface can have important consequences for the derived constraints.
Nath, Strohmayer \& Swank (2001) have modelled bolometric pulse
profiles observed during the rising phase of bursts from 4U
1636-53. They fit the pulse profiles with a rotating, expanding hot
spot model that includes light deflection in the Schwarzschild
spacetime. They find that the inferred constraints depend very
sensitively on whether or not two spots are present.  Much more
restrictive compactness constraints can be achieved if two spots are
present. The main reason being that large amplitudes are much more
difficult to achieve with two spots than one. Assuming two hot spots
they find a lower limit to the compactness of $m/r < 0.163$ at 90\%
confidence. This requires a relatively stiff equation of state for the
neutron star interior and disagrees with the recently measured
value of $m/r=0.23$ from Cottam et al. (2002). If one hot spot is assumed,
then the constraint is consistent with the Cottam et al. (2002) measurement.
Weinberg, Miller \& Lamb (2001) have explored
the oscillation waveforms and amplitudes produced by rotating neutron
stars with single and antipodal hot spots of varying size.  They
include photon propagation in the Schwarzschild spacetime and consider
the effects of relativistic aberration and Doppler shifts induced by
the rotational motion of the neutron star surface.  They concluded
that pulse profile fitting could be a powerful tool in constraining
neutron star properties.

\subsection{Spin modulation: implications for spin in LMXBs}

The discovery of burst oscillations provided the first strong
observational evidence for millisecond rotation periods in accreting
neutron star LMXB's.  Since then three bonafide accreting millisecond
pulsars have been discovered; SAX J1808.4--3658 (Wijnands \& van der
Klis 1998; Chakrabarty \& Morgan 1998), XTE J1751--305 (Markwardt et
al. 2002), and XTE J0929--314 (Galloway et al. 2002), with spin
frequencies of 401 Hz, 435 Hz, and 185 Hz, respectively. These pulsars
all have extremely low mass companions, with $M_c \approx 0.01
M_{\odot}$ (for XTE J1751--305 and XTE J0929--314, Bildsten 2002), and they 
have firmly established the link between millisecond radio pulsars and
accreting LMXBs. However, until very recently no source had shown both
persistent pulsations and burst oscillations \textit{at the same
frequency}, with the exception of a $3\sigma$ detection with the WFC
of an oscillation at 401 Hz during a burst from SAX J1808.4--3658 (in
't Zand et al.  2001).

A few weeks before having to finalize our review, a new outburst of
the accreting millisecond pulsar SAX J1808.4--3658 was discovered with
the RXTE/ASM (Markwardt, Miller \& Wijnands 2002). Extensive RXTE
observations of the outburst have so far detected four thermonuclear
bursts, all of which are PRE bursts.  Each burst also shows
oscillations, with pulsations detected during the rise and after
``touchdown'' of the photosphere, as is seen in many other burst
oscillation sources which do not show pulsations in their persistent
emission. The oscillations in the bursts from SAX J1808.4--3658 are
clearly distinct from the coherent pulsations in the persistent flux
because the oscillation amplitude is larger. Oscillations seen in the
decaying tails were at the spin frequency (Wijnands et al. 2002). These
recent findings, as well as the discovery of coherent pulsations
during a superburst from 4U 1636--53 (see Strohmayer \& Markwardt 2002;
\S 3.5.4 below) conclusively establish that burst oscillations result
from spin modulation of the X-ray burst flux.

\subsection{Theoretical implications of burst oscillations} 

The spin modulation mechanism requires a slow-moving, non-uniform
brightness pattern on the neutron star surface.  At burst onset the
pattern is most likely a localized ``hot spot'', whereas we are still
mostly in the dark as to the origin of oscillations long after the
burning has started.  What all of the observations have made clear
is that we can no longer persist with spherically symmetric modelling.
In many ways, the question we now need to answer is: ``What breaks the
symmetry?'' as prior to and after the bursts, there is no
indication of azimuthal variations on the stellar surface.

\begin{figure*}[tbp]
 \centering 
\vspace{1cm}
 \epsfig{file=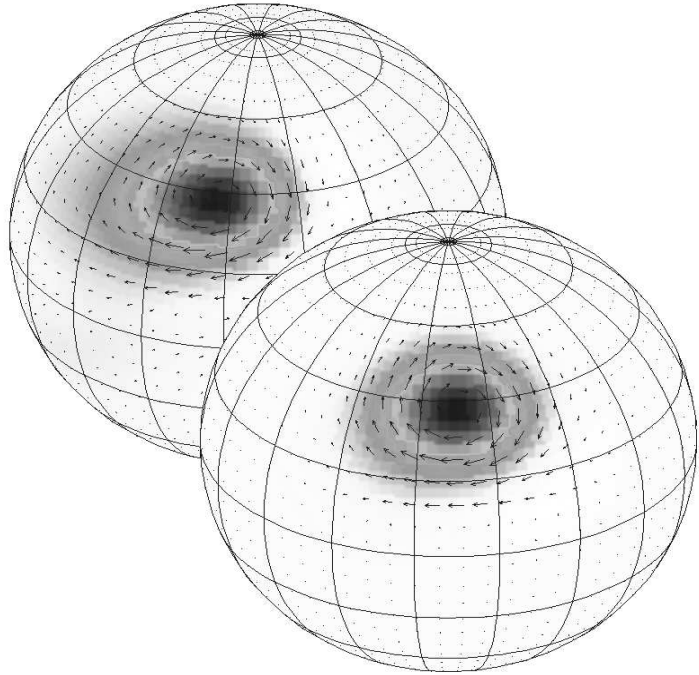,height=3.75in,width=4.25in} 
\caption{Initial evolution of a burning hot spot ignited off the
equator as seen in a frame rotating with the neutron star. Velocity
vectors show the circulation of the fluid induced by the Coriolis
forces. The hot spot expands due to burning and drifts west-southwest
because of the latitude dependence of the Coriolis force (after
Spitkovsky, Levin \& Ushomirsky 2002).}  
\label{figure13}
\end{figure*}

The initial work on the spreading of a locally ignited (i.e. symmetry
broken by hand) ``hot spot'' away from the ignition site were focused
on the laminar or convective combustion viewpoint (e.g. Fryxell \&
Woosley 1982; Bildsten 1995) and neglected the physics of the
atmospheric response to transverse pressure gradients. 
Such an approximation is only appropriate if the matter is very
degenerate. In this limit, the observed short rise times can only be 
explained with convective velocities. 
However, for most observed bursts the degeneracy is only
partial during the flash. In this case, an additional velocity becomes
important, which is the shallow water wave speed, $V_{sw}\approx
(gh)^{1/2}$, where $g$ is the surface gravity and $h\approx 10 $
meters is the thickness of the burning layer. This speed is $\approx
4000$ km/sec, so that, in the absence of rotation, any transverse
pressure disturbances create wave-like disturbances with periods as
short as 5 -- 10 milliseconds (e.g. Livio \& Bath 1982; McDermott \& Taam 1987;
Bildsten \& Cutler 1995; Bildsten \& Cumming 1998). However, these periods are
longer than the neutron star rotation period, in which case the
Coriolis force must be taken into account (see Bildsten et al. 1996;
Strohmayer \& Lee 1996 for the rotational modification of modes) when
considering the nature of the modes.

Spitkovsky et al. (2002) showed that this same interplay between the
shallow-wave speed and the Coriolis force is relevant to the
propagation speed of localized burning on a rotating star. In their
groundbreaking calculation, they found that the width of the burning
front at the leading edge of the hot spot was the Rossby adjustment
radius $\sim V_{sw}/\Omega$, and the resulting speed of the front is
$V_f\sim V_{sw}/(\Omega t_n)\sim 2-10 \ {\rm km \ s^{-1}}$, where
$t_n\sim 0.1-1 $ second is the time to burn the fuel (for example, in
a helium-rich flash). The spreading time for the whole star is then
0.1 to a few seconds, in the observed range. The dependence on the
rotation rate, $\Omega$, would be nice to test, but the current range
of measured rotation rates in bursters is likely too small to allow
for it. 
   
Spitkovsky et al. (2002) also investigated the dynamics of the nuclear
ignition on a rotating star and argued that ignition will tend to
occur in the equatorial region and propagate to the poles (see Figure
3.13). The combination of radial uplift and horizontal flows they found
may also be able to explain the observed frequency drifts. This
initial progress is remarkable, but much remains to be done, including
an implementation of more realistic burning during the propagation and
an improved understanding of where the accreted fuel resides and
where/how ignition really starts in a three-dimensional star.  The
standard ignition condition is for a spherically symmetric
perturbation of a spherically symmetric model and most likely both of
those approximations need to be dropped.

In the burst tails the oscillation might conceivably be produced by a
mode (e.g. McDermott \& Taam 1987) or perhaps is generated dynamically
by the interaction of burst heating and cooling with the rapid spin of
the star (see Spitkovsky et al. 2002). Indeed, the pure harmonic
content points to an azimuthal perturbation $\propto \exp
(im\phi)$. However, any such model must explain both the $\approx 1
\%$ frequency drift and the long-term stability and none convincingly
have.

Without specifically addressing the mode question, Cumming \& Bildsten
(2000) explored in detail the radial uplift mechanism outlined by
Strohmayer et al. (1997) to explain the frequency drifts. In the
context of presuming that the burning layer became disconnected from
the underlying material but itself rigidly rotated, they concluded
that this process could explain most of the drifts. Heyl (2000)
claimed that properly including general relativistic effects would
allow for strong constraints on the NS radius, but soon thereafter
Abramowicz, Kluzniak \& Lasota (2001) and Cumming et al. (2002) found
an error in Heyl's formulation that strongly reduced the impact of
general relativity.  However, during this reevaluation, Cumming et
al. (2002) found that they had overestimated the radial uplift drift
by about a factor of two.  This result, combined with the observations
of much larger drifts in several sources (Galloway et al. 2001,
Wijnands et al. 2002) now makes it appear that radial uplift is
insufficient to account for all the observed drift if the whole
burning layer is rigidly rotating. One possible way out from this
conundrum is to allow some fluid elements in the burning layer to keep
their initial angular momentum (i.e. not demand rigid rotation of the
whole burning layer).  This easily produces much larger frequency
shifts for the outermost layers (Cumming and Bildsten 2000), but the
origin of such a coherent signal in the context of the rapid internal
differential rotation is then a new puzzle to solve.  These
uncertainties will likely remain with us until the origin of the
asymmetry is resolved.

The strength of the large-scale dipole field that threads the burning
layers on the neutron star is unknown. The lack of persistent
pulsations during accretion leads to the limit of $B<10^{8-9} $ G by
presuming that the spherical magnetospheric radius is inside the
neutron star. However, $B$ might need to be even lower than this to
ensure azimuthal symmetry. For example, it is easy to imagine that a
permanent asymmetry could result even if the material arrives on the
equator, but encounters an ordered field as it tries to spread away as
the photospheric pressure is only $\approx 10^{15} {\rm erg \
cm^{-3}}$. In addition, during the bursts, if the frequency drift that
is observed is due to vertical shear from the radially expanded
burning layers, an initially poloidal field as weak as $10^6 \ {\rm
G}$ could become dynamically important as it is wound up by
differential rotation (Cumming \& Bildsten 2000).

\section{Superbursts: a new burning regime}

Since the advent of BeppoSAX and RXTE in 1996, the X-ray sky has been
monitored with unprecedented sensitivity and frequency.  This
capability has opened up the discovery space for burst events with
long (years) recurrence times, which were apparently missed by
previous missions. Cornelisse et al. (2000) reported the first
superburst discovery (from BeppoSAX) in the familiar Type I burster 4U
1735--44. This was rapidly followed by reports of six more long X-ray
flares lasting 3 - 5 hours from five previously known X-ray bursters;
4U 1820--30 (Strohmayer 2000, Strohmayer \& Brown 2001); KS 1731--260
(Kuulkers et al. 2002c); 4U 1636--53 (Wijnands 2001; Strohmayer \&
Markwardt 2002); Ser X-1 (Cornelisse et al. 2002c); and GX 3+1
(Kuulkers 2002). One of the sources, 4U 1636--53, produced two bursts
separated by 4.7 years (Wijnands 2001). Two of the events, one from 4U
1636--53, and the other from 4U 1820--30 were observed with the large
area PCA onboard RXTE (Figures 3.14 and 3.15 show superbursts seen 
with the BeppoSAX/WFC and RXTE/PCA, respectively).

\begin{figure*}[tbp]
 \centering \vspace{0cm}
 \epsfig{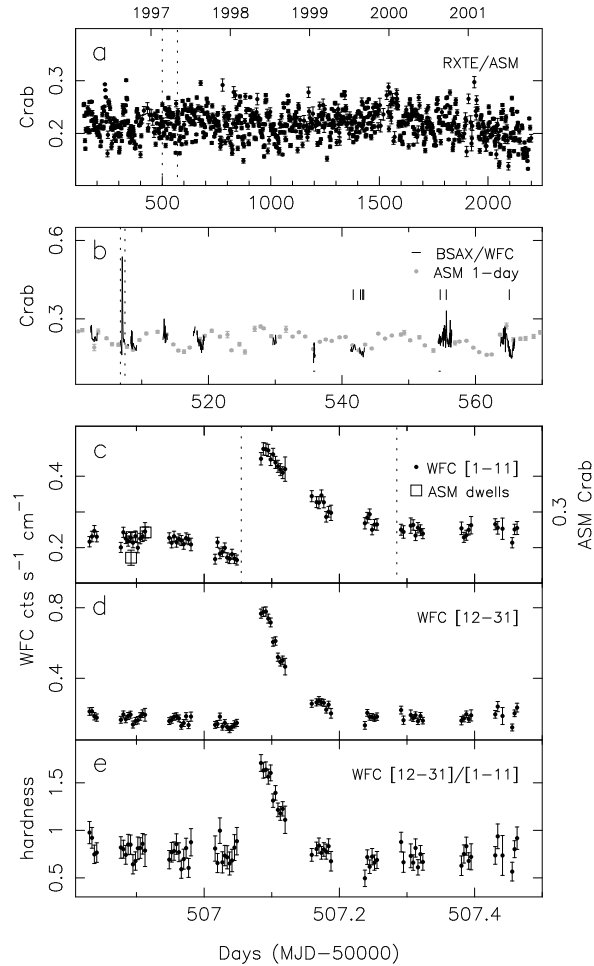} 
\caption{The superburst from Ser X-1 observed with the WFC on BeppoSAX. 
Note the persistent offset in the flux after the superburst in
the 2 - 5 keV band (c). The hardness ratio (e) drops through the burst,
indicative of cooling of the neutron star surface (after Cornelisse et al. 
2002c).}  
\label{figure14}
\end{figure*}

Table 3.2 (after Kuulkers et al. 2002c) summarizes the important
properties of the superbursts observed to date.  These flares show all
the hallmarks of thermonuclear bursts; they have thermal spectra which
soften with time, they show more or less smooth exponential-like
decays, and some show PRE episodes. The fundamental distinctions
between these events and standard type I X-ray bursts are their long
durations, larger fluences and long recurrence times (ie. they are
rare).  Indeed, these events are typically 1,000 times longer and more
energetic than standard X-ray bursts, because of this, they have come
to be referred to as ``superbursts,'' and we will continue to use this
appellation. Although the sample size is still relatively small, the
systems which have produced superbursts have typical accretion rates
in the range $\approx 0.1 - 0.3$ of $\dot M_{Edd}$ (Wijnands 2001;
Kuulkers et al. 2002c). As we describe in more detail shortly, this is
an important constraint on theoretical models.

The durations and energetics of superbursts suggest that they result
from thermonuclear flashes occurring in fuel layers at much greater
depth than for typical X-ray bursts. For example, if the nuclear
energy release is 0.3 MeV per accreted nucleon, the accumulated mass
required to power a $10^{42} {\rm erg}$ superburst is $3.5\times
10^{24} \ {\rm g}$, giving a recurrence time of $\approx 2$ years for
an accretion rate of $\dot M\approx 10^{-9} M_\odot \ {\rm yr^{-1}}$.
In the remainder of this section we will summarize the current state
of knowledge of superbursts and explore what new insights they are
giving us about nuclear burning on neutron stars.

\subsection{Time profiles and spectra}

The most distinctive characteristic of superbursts are their long
durations.  The time profiles of most superbursts have been
characterized with an exponential decay time, and these values range
from about 1 - 3 hours (see Kuulkers et al. 2002c). The longest event
seen to date was the superburst from KS 1731--260, which lasted for
almost half a day (Kuulkers et al. 2002c).  Their time profiles are
generally smooth, with a fast rise and exponential decay, however, not
all superbursts have been observed with comparable sensitivity and
temporal resolution, so comparisons must be made with some
caution. For example, the superburst from 4U 1820--30 showed
substantial variability during portions of the decay phase (Strohmayer
\& Brown 2002).  Both superbursts which were observed with the large
area PCA show ``precursor'' events just prior to the start of the
superburst (see Strohmayer \& Brown 2002; Strohmayer \& Markwardt
2002; Figure 3.15). These precursors look more or less like standard 
type I bursts from the respective objects.  There is also some evidence for a
precursor in the burst from KS 1731--260 observed with the BeppoSAX/WFC
(Kuulkers et al. 2002c).

Based on the available data, it appears that such
precursors are a common feature of superbursts. If the superbursts
result from energy release at great depths, then it seems plausible
that the superburst flux could trigger a flash in the H/He layers
above it as it diffuses outward. There has been some speculation that
the precursor may act as a trigger for the superburst, but given the
much longer radiative diffusion time at the depth where the
superbursts are likely triggered, this seems unlikely (see Strohmayer
\& Brown 2002).

\begin{table*}[tbp]
\begin{center}
\footnotesize
 \caption{Superburst sources and properties (after Kuulkers et al. 
2002c)}
  \begin{tabular}{lcccccc}
   \hline \hline
Source & 4U 1820--30 & 4U 1735--44 & KS 1731--260 & 4U 1636--53 & Ser X-1 & 
GX 3+1 \\
detector$^a$ & PCA & WFC & WFC, ASM & PCA, ASM & WFC & ASM \\
Duration$^b$ & 3 & 7 & 12 & $>2-3$ & $\approx 4$ & $> 3.3$ \\
Precursor? & yes & ? & yes & yes & ? & ? \\
$\tau_{exp} \rm{(hr)}$ & $\approx 1$ & $1.4\pm 0.1$ & $2.7 \pm 0.1$ & 
$1.5 \pm 0.1$ & $1.2 \pm 0.1$ & $1.6 \pm 0.2$ \\
 & & & & $3.1 \pm 0.5$ & & \\
$L_{pers}$$^{c}$ & $\approx 0.1$ & $\approx 0.25$ & 
$\approx 0.1$ & $\approx 0.1$ & $\approx 0.2$ & $\approx 0.2$ \\
$kT_{max}$$^d$ & $\approx 3$ & $\approx 2.6$ & $\approx 2.4$ & ? & $\approx 
2.6$ & $\sim 2$ \\
$L_{peak}$$^e$ & 3.4 & 1.5 & 1.4 & 1.2 & 1.6 & 0.8 \\
$E_{b}$$^f$  & $> 1.4$ & $> 0.5$ & $\approx 1$ & $ 0.5 - 1$ & 
$\approx 0.8$ & $> 0.6$ \\
$t_{quench}$$^g$ & ? & $> 7.5 $ & $> 35$ & ? & $\sim 34$ & ? \\
References$^h$ & S00, SB02 & C00 & K02c & W01, SM02 & C02 & K02 \\
\hline \hline
\end{tabular}

\end{center}
\begin{minipage}{0.95\linewidth}
$^a$ Instruments which observed the superburst.

$^b$ In hours.

$^c$ Persistent luminosity prior to superburst, in terms of the Eddington 
luminosity. 

$^d$ In keV.

$^e$ In units of $10^{38}$ ergs s$^{-1}$.

$^f$ in units of $10^{42}$ ergs.

$^g$ Time following superburst with no normal burst activity, in days.

$^h$ S00 (Strohmayer 2000); SB02 (Strohmayer \& Brown 2002); C00 
(Cornelisse et al. 2000); K02c (Kuulkers et al. 2002c); W01 (Wijnands 2001); 
SM02 (Strohmayer \& Markwardt 2002); C02 (Cornelisse et al. 2002c); K02
(Kuulkers 2002).

\end{minipage}

\label{table2}
\end{table*}
\normalsize

The X-ray spectra of superbursts are well described by thermal
emission, with peak blackbody temperatures in the range from 2 -- 3
keV, quite typical for thermonuclear bursts (Cornelisse et
al. 2000). Interestingly, most of the observed superbursts have peak
fluxes which are sub-Eddington, only the superburst from the pure
helium accretor 4U 1820--30, had a peak flux consistent with the
Eddington limit.

Strohmayer \& Brown (2002) found discrete
components in the spectra of the superburst from 4U 1820--30; a broad
emission line centered near 6 keV and an accompanying absorption edge
between 8 and 9 keV (see Figure 3.5). They suggested that these features 
can be explained by reflection of the burst flux from the inner accretion
disk around the neutron star. Day \& Done (1991) had predicted that
such features might be detectable in spectra of bursts. The energy and
width of the dominant Fe K$\alpha$ fluorescence line as well as the
energy of the edge can be used as important diagnostics of the
ionization state of the disk (see for example, Ross, Fabian \& Young
1999; Nayakshin \& Kallman 2001). An origin of the features in a
burst-driven wind is also possible. Because superbursts last about a
1,000 times longer than normal bursts and can give very high signal to
noise spectra, they would allow much more sensitive searches for
discrete spectral lines from neutron star surfaces.  Superbursts would
therefore make tempting targets for rapidly triggered observations
with sensitive, high spectral resolution observatories. The soon to be
launched Swift mission may be able to provide the necessary triggering
capability.

\subsection{Superburst energetics}

The available X-ray spectroscopy on superbursts provides estimates of
the total X-ray energy liberated by the events.  For the superburst
from 4U 1820--30 a lower limit to the energy fluence was $1.5 \times
10^{42}$ ergs, assuming a distance of 6.6 kpc (Strohmayer \& Brown
2002). It is extremely unlikely that unstable helium burning could
provide such a large fluence. This could only occur at very low
atmospheric temperatures and thus at low mass accretion rates
inconsistent with the persistent X-ray flux observed from 4U 1820--30
(Fryxell \& Woosley 1982; Zingale et al. 2001).  Similar fluence
limits have been derived for the other superbursts as well (see
Kuulkers et al. 2002c).  The total energy observed in X-rays, combined
with the decay timescales of superbursts strongly argues for a fuel
source located at depths below the column density where helium flashes
are triggered.

\subsection{Quenching of normal burst activity}

There are strong indications from several of the superburst sources
that the occurrence of a superburst has a profound influence on the
thermal state of the accreted ``ocean'' on the neutron star.  In
particular the occurrence of normal (short duration) X-ray bursts
appears to be suppressed for some time following superbursts.  For
example, Kuulkers et al. (2002c) and Cornelisse et al. (2002c) found
that normal bursting ceased for about 35 days following the
superbursts from KS 1731--260 and Ser X-1, respectively (see Figure
3.14, where the tick marks in panel b denote the positions of
normal bursts). There are also indications that normal bursting was
suppressed for at least a week following the superburst from 4U
1735--44 (Cornelisse et al. 2000). A likely explanation of this
suppression is that flux from the deep parts of the neutron star ocean
remains high enough to quench the Type I bursting activity (Cumming \&
Bildsten 2001).

\begin{figure*}[tbp]
 \centering \vspace{0.25cm}
\epsfig{file=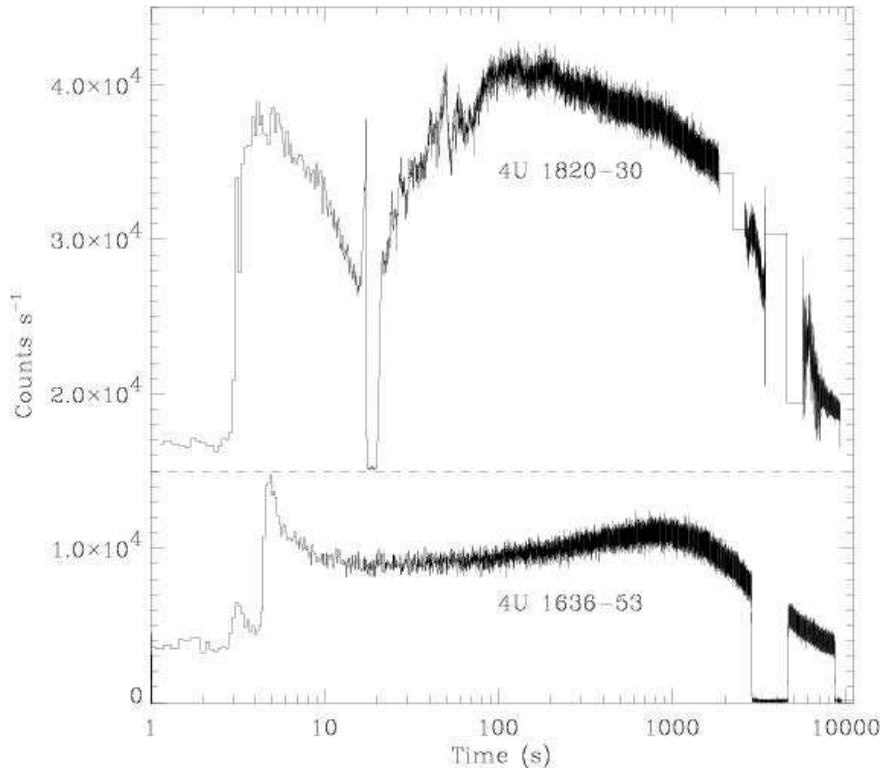,width=0.9\linewidth}
\caption{Two superbursts observed with the RXTE/PCA. Shown are the 
2 - 30 keV count rate histories observed in the PCA. Note the shorter 
precursor events prior to the superbursts. The event from 4U 1820--30 has
been displaced vertically for clarity.  The horizontal dashed line shows the
zero level for this event. The time axis is logarithmic 
(after Strohmayer \& Brown 2002; Strohmayer \& Markwardt 2002).}
\label{figure15}
\end{figure*}

\subsection{Millisecond pulsations during a superburst: 4U 1636--53}

Burst oscillations at $\approx 582$ Hz have been observed in many
bursts from the LMXB 4U 1636--53 (see Table 3.1; Zhang et al. 1997:
Giles et al. 2002).  Strohmayer \& Markwardt (2002) recently
discovered coherent pulsations at this frequency during a portion of
the February 22, 2001 (UT) superburst from this source. They were
detected during an $\approx 800$ s interval spanning the flux maximum
of the superburst. The average pulsation amplitude was $1 \%$, which
is smaller than the amplitudes of oscillations observed from standard
bursts. The pulse trains observed during the superburst are much
longer than the typical, 10 s long pulse trains observed in normal
bursts. The pulsation frequency was found to increase in a monotonic
fashion by less than a part in $10^4$ (see Figure 3.16). This is a
much smaller frequency drift than commonly seen in burst oscillations
from standard bursts (see \S 3.4.4). The form of the frequency
evolution appears consistent with Doppler modulation caused by the
known orbital motion of the neutron star around the center of mass of
the binary.  Strohmayer \& Markwardt (2002) showed that a circular
orbit model fits the observed frequency evolution well. The best phase
evolution model is consistent with a coherent pulsation during the
observation interval, and gives a limit on the coherence $Q \equiv
\nu_0 / \Delta_{\nu_0} > 4.5 \times 10^5$.  The orbital fits indicate
that the projected neutron star velocity lies between 90 and 175 km
s$^{-1}$. The brevity of the observed pulse train with respect to the 
orbital period of 3.8 hr does not allow for more precise constraints.
The fact that the coherent pulsation frequency during the
superburst is within $\approx 1$ Hz of all the measured asymptotic
burst oscillation frequencies for 4U 1636--53 indicates that the
frequencies are set by the spin of the neutron star.

\begin{figure*}
 \centering 
 \epsfig{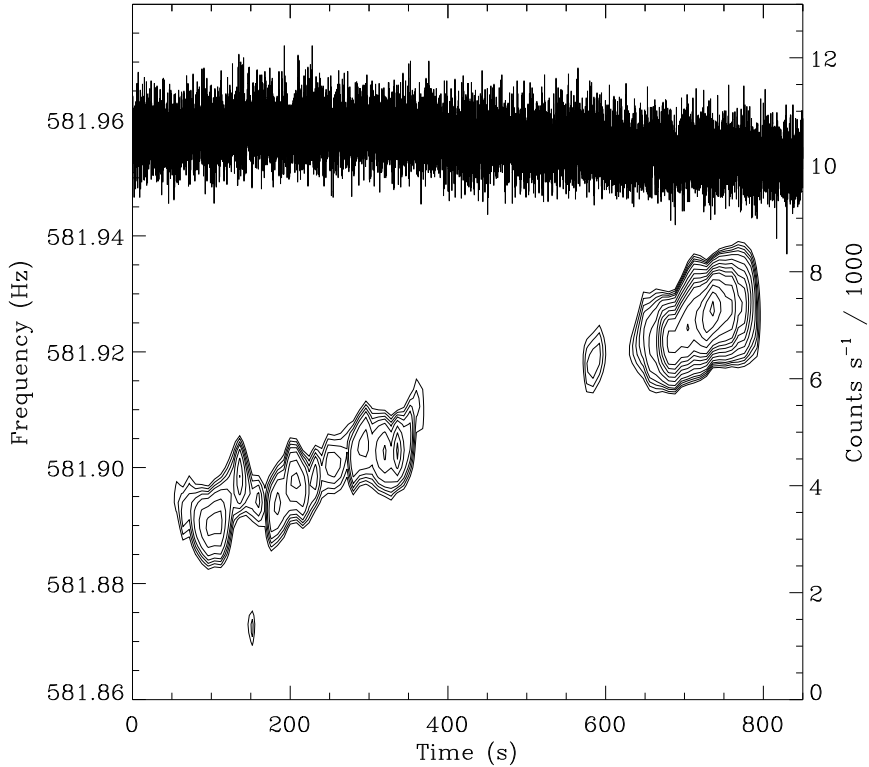}
 \caption{Dynamic power spectrum of a portion of the February 22, 2001
 superburst from 4U 1636--53 showing 582 Hz pulsations. Shown are
 contours of constant $Z_1^2$ power versus frequency (left axis), and
 the RXTE/PCA countrate (right axis), both as a function of time. The 
monotonic increase in the pulsation frequency is consistent with binary 
orbital modulation (after Strohmayer \& Markwardt 2002).}  
\label{figure16}
\end{figure*}

\subsection{Theory of Superbursts: Ashes to Ashes}

Unstable burning of a pure carbon layer (Woosley \& Taam 1976; Taam \&
Picklum 1978; Brown \& Bildsten 1998) has been proposed as an
explanation for the superburst from the pure helium accretor 4U
1820--30 (Strohmayer \& Brown 2002). In this scenario, the accumulated
mass of carbon at ignition is $10^{26}-10^{27}\ {\rm g}$, giving
decades long recurrence times and most of the energy released
($10^{43}$--$10^{44}\ {\rm ergs}$) escapes as neutrinos or is
conducted into the star, leaving $\approx 10^{42}\ {\rm ergs}$ to
emerge from the surface within a few hours (Strohmayer \& Brown
2002). Shorter recurrence times are possible if a smaller mass of
carbon can be triggered somehow.  Large carbon fractions in the 
ocean are expected for 4U 1820--30, since stable helium burning at
the higher inferred mass accretion rates, when normal (10 - 20 second 
duration) bursts are not observed, will produce lots of carbon.

Pure carbon is unlikely to apply to
the superbursts from H/He accretors as Schatz et al.~(1999, 2001) have
shown that only a small amount of carbon remains after the burning of
H and He via the rp-process.  However, Cumming \& Bildsten (2001)
(hereafter CB01) showed that even small amounts of carbon can be a
promising energy source for the superbursts. They found that burning
of this small mass fraction of carbon is thermally unstable at low
accumulated masses when the ocean contains heavy ashes from the
rp-process.

CB01 proved the important role played by the rp-process ashes.  Their
low thermal conductivity gives a large temperature gradient in the
ocean, so that the trace carbon ignites at accumulated masses
comparable to that observed. The resulting energies, recurrence times,
and conductive cooling times can easily accommodate the observed
properties of superbursts (see Figure 3.17), especially now that an
extra energy source was found by Schatz, Bildsten \& Cumming (2003). 
They showed that the conversion of rp-process nuclei back to the iron
group elements during the carbon-triggered flash can enhance the
energy release from that given in CB01 by factors of four. 

The instability requires 
that $\dot M > 0.1 \dot M_{\rm Edd}$ when the carbon mass fraction is 
less than 10\%. Lower $\dot M$'s stably burn the carbon. Though the 
instability is present at accretion rates $\approx \dot M_{\rm Edd}$, 
those flashes provide less of a contrast with the accretion luminosity, 
thus explaining why detection is easier when $\dot M\approx (0.1-0.3)\dot
M_{\rm Edd}$. Detecting one of these flares from a rapidly accreting Z
source requires flux sensitivity at the $10\%$ level on a timescale of
a few hours and spectral sensitivity to distinguish that the flux rise
is from extra thermal emission. This should be carried out and would
confirm the notion of trace carbon ignition in the heavy rp-process
ashes.

\begin{figure*}[tbp]
 \centering 
 \epsfig{file=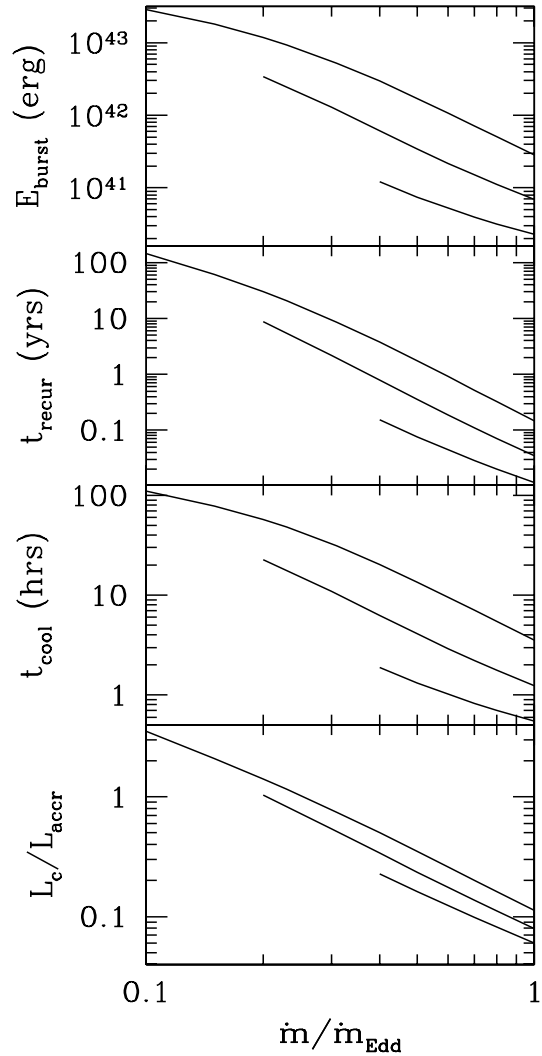,height=5.5in,width=0.55\linewidth}
 \caption{Results of theoretical calculations of carbon flashes in the heavy 
element ocean on a neutron star. Shown from top to bottom are the burst 
energies, recurrence times, luminosity decay times, and the ratio of cooling
luminosity to accretion luminosity, as a function of $\dot m$. The curves 
begin at the lowest accretion rate at which the thermal instability occurs 
in the heavy element ($^{104}$Ru) ocean. (after Cumming \& Bildsten 2001).}  
\label{figure17}
\end{figure*}

The energy from these mixed flashes takes a long time to escape the
star, possibly explaining the persistent ``offset'' in the flux nearly
a day after the superburst in Ser X-1 (see Figure 3.14). CB01 also
attributed the halting of regular Type I bursts after the superbursts
to thermal stabilization of the H/He burning layers by the large heat
flux from the cooling ashes of the carbon burning.  Paczynski (1983)
and Bildsten (1995) have shown that luminosities in excess of the
helium burning flux (or $L>L_{\rm accr}/100$) will stabilize the H/He
burning. This thus provides additional evidence that the superbursts
are from burning deep within the neutron star.

An alternative to carbon scenarios has been discussed by Kuulkers et
al. (2002c).  They suggest that electron captures on protons might be
able to supply sufficient energy to power superbursts.  The Fermi
energy at column depths required to account for superburst energetics
is close to the proton - neutron mass difference, so that protons can
easily capture electrons. The resulting neutrons can then be captured
on heavy nuclei, supplying about 7 MeV per nucleon (Bildsten \&
Cumming 1998). It is not yet clear whether this process is thermally
unstable under the relevant conditions, and it is not certain that
sufficient hydrogen can survive the initial hydrogen/helium burning.
Further theoretical work should be able to answer these questions.

The initial theoretical understanding of superbursts is an exciting
development as it further motivates the nucleosynthesis studies during
Type I bursts and connects the ashes from the rp-process burning to
the superburst explosions. We also hope that future observations will
find the equivalent of superbursts from the higher accretion rate Z
sources, or even from accreting X-ray pulsars.

\section{Summary and Future Prospects}

There has clearly been tremendous growth in our observational
understanding of thermonuclear bursts in the past decade.  Indeed, we
think it is fair to say that the observations have gotten
significantly ahead of the theory at the time of this writing. Yet, the
promise of probing neutron star structure, and the fundamental physics
needed to describe it, with burst observations has never been greater.  
Our understanding, though still incomplete, of the new phenomena, as well 
as insights drawn from previous observations, are sufficient to show that
researchers must really begin to explore fully three dimensional and
time dependent calculations of nuclear energy production and radiation
transport in the neutron star surface layers in order to fully exploit
the new phenomena as probes of neutron star structure and fundamental
physics. The days of spherically symmetric, static models being
adequate are long gone in our opinion.

In the context of burst oscillations it seems clear that the spreading
of the nuclear burning front has important observational consequences,
particularly for oscillations seen near the onset of bursts.  A
thorough theoretical understanding will only come with
multi-dimensional dynamical studies of nuclear ignition and
propagation. Indeed, higher signal to noise observations of the
oscillations during burst rise could in principle provide us with a
snapshot view of how burning spreads. Such studies will require even
larger X-ray collecting areas than RXTE. A future goal would be to
have detailed theoretical models of front propagation available for
comparison with new data by the time that such observations are
eventually made. Moreover, the persistence of oscillations in the
cooling phase suggests that the dynamics of burst heating and rapid
rotation combine to produce quasi-stable patterns in the surface
layers that can persist for thousands of rotation periods (Spitkovsky
et al. 2002).  It is a triumph of modern X-ray astronomy that we are
able to infer changes in the surface layers on neutron stars on the
scale of only 10's of meters at distances of kiloparsecs! 
Helioseismology has revolutionized study of the solar interior,
perhaps future improvements in sensitivity will allow a similar
revolution in neutron star studies by enabling the detection of global
oscillation modes of neutron stars.

As we have shown, an understanding of superbursts likely requires
knowledge of the by-products of the nuclear burning. Detailed models
of the heat and radiation transport from the deep ocean to the surface
will be required to accurately model surface fluxes and time profiles
and should be able to inform us about the location and conditions
where superbursts are triggered.  The existing RXTE observations with
high signal to noise but low spectral resolution have shown that
superbursts are likely promising targets for higher resolution
spectroscopy, one of the goals of which should be to detect line
features from neutron star surfaces. If such observations are to be
possible, then X-ray monitoring instruments must be in place which can
generate the necessary triggers.  The discovery of superbursts was
made possible with intensive all sky X-ray monitoring. Future efforts
to study superbursts, for example, to better constrain the recurrence
times, will require continued and improved X-ray monitoring
capabilities. Future missions which should be able to address these
needs include, INTEGRAL, Swift, Lobster-ISS, and MAXI.

High spectral resolution studies of bursts have begun with the new
capabilities of Chandra and XMM/Newton. Frustratingly, the present
data suggests that isolated neutron stars may be poor targets for line
searches perhaps because of a dearth of metals in the
atmosphere. Accreting neutron stars could be more tempting targets,
with a continuous supply of metals furnished by the mass
donor. Although the new generation of X-ray observatories have
impressive capabilities, they still lack the collecting area necessary
to obtain high signal to noise spectra from single bursts. To do
better will require larger area missions, such as NASA's
Constellation-X and ESA's XEUS.

We thank Erik Kuulkers, Craig Markwardt, Deepto Chakrabarty, Hendrik Schatz 
and Jean Swank for comments on the manuscript. We are indebted to Mike Muno, 
Andrew Cumming, Jean Cottam, Lucia Franco, Hendrik Schatz, Duncan Galloway, 
Remon Cornelisse, Erik Kuulkers and Anatoly Spitkovsky for either providing 
graphics or allowing us the use of previously published figures. We sincerely 
thank them all. We would like to dedicate this work to the memory of John 
C. L. Wang, a friend and colleague who left us much too soon. This work was 
supported by the National Science Foundation under grants PHY99-07949 and 
AST02-05956, NASA through grant NAG 5-8658.  L. B. is a Cottrell Scholar of 
the Research Corporation.

\end{document}